\documentclass[12pt,english, usletter]{article}
\usepackage{lmodern}

\usepackage[OMLmathrm,OMLmathbf]{isomath} 

\usepackage{nicefrac}
\usepackage[T1]{fontenc}
\usepackage[latin9]{inputenc}
\usepackage[letterpaper]{geometry}
\usepackage[section]{placeins}
\usepackage[]{caption}
\usepackage{bbm}
\usepackage{amsmath}
\usepackage{amsthm}
\usepackage{amssymb}
\usepackage{setspace}
\usepackage[authoryear]{natbib}
\usepackage{calrsfs}

\usepackage{mathrsfs}
\usepackage[mathcal]{euscript}
\DeclareMathAlphabet{\pazocal}{OMS}{zplm}{m}{n}
\usepackage{xcolor}
\usepackage[colorlinks,linkcolor=links,citecolor=cites,urlcolor=MyDarkBlue]{hyperref}
\definecolor{MyDarkBlue}{rgb}{0,0.08,0.45}
\definecolor{cites}{HTML}{324b13}
\definecolor{links}{HTML}{1a663b}
\definecolor{MyLightMagenta}{cmyk}{0.1,0.8,0,0.1}

\makeatletter
\theoremstyle{plain}
\usepackage{babel}
\usepackage{tikz}
\usepackage{pgfplots}
\usepgfplotslibrary{fillbetween}
\usepgfplotslibrary{patchplots}
\usepackage{subfig}

\def\lemmaname{Lemma}
\def\propositionname{Proposition}

\def\assumptionname{Assumption}
\def\corollaryname{Corollary}

\usepackage[multiple]{footmisc} 
 \tikzset{ every node/.style={inner sep=0pt,minimum size=1mm},
  nsnode/.style={draw,circle,black},
    nsnode2/.style={draw,circle,blue},
  nnnode/.style={draw,circle,black,fill=black},
  asnode/.style={draw,circle,blue},
  bsnode/.style={draw,circle,green,fill=green},
  csnode/.style={draw,circle,red,fill=red, minimum size=2mm},
  every fit/.style={inner sep=-1.5pt,text width=1cm}  }
  \tikzset{nero/.style={decorate,draw=black}}
\tikzset{bianco/.style={decorate,draw=bg}}

 \newcommand*\dotp{\mathpalette\dotp@{.5}}
\newcommand*\dotp@[2]{\mathbin{\vcenter{\hbox{\scalebox{#2}{$\m@th#1\bullet$}}}}}

\newcommand{\prior}{\ensuremath{\mu_0}}
\newcommand{\state}{\ensuremath{\omega}}
\newcommand{\states}{\ensuremath{\Omega}}
\newcommand{\action}{\ensuremath{a}}

\newcommand{\amin}{\ensuremath{\underline{a}}}
\newcommand{\amax}{\ensuremath{\overline{a}}}
\newcommand{\atotal}{\ensuremath{\breve{a}}}
\newcommand{\atotali}{\ensuremath{\breve{a}_{-i}}}

\newcommand{\actions}{\ensuremath{A}}
\newcommand{\reals}{\ensuremath{\mathbb{R}}}
\newcommand{\var}{\ensuremath{\sigma^2}}
\newcommand{\up}{\ensuremath{v}}
\newcommand{\ui}{\ensuremath{u_i}}
\newcommand{\ud}{\ensuremath{w}}
\newcommand{\udl}{\ensuremath{W}}
\newcommand{\Bp}{\ensuremath{B}}
\newcommand{\Cp}{\ensuremath{C}}

\newcommand{\belief}{\ensuremath{\mu}}
\newcommand{\expect}{\ensuremath{\mathbb{E}}}

\newcommand{\resp}{\ensuremath{R}}

\newcommand{\info}{\ensuremath{\pazocal{I}}}

\newcommand{\q}{\ensuremath{q}}

\newcommand{\uconsmat}{\ensuremath{H}}

\newcommand{\sensni}{\ensuremath{\eta}}
\newcommand{\cs}{\ensuremath{CS}}
\newcommand{\profits}{\ensuremath{PS}}

\newcommand{\mean}{\ensuremath{\omega_0}}

\newcommand{\argmin}{\operatorname{arg\,min}}

\newcommand{\deltacr}{\ensuremath{\hat\delta}}
\newcommand{\deltaFB}{\ensuremath{\overline{\delta}}}
\newcommand{\ri}{\ensuremath{r_o}}
\newcommand{\rni}{\ensuremath{r_c}}

\newcommand{\VAR}{\ensuremath{\mathbb{V}}}

\newcommand{\meas}{\ensuremath{\mathcal{M}}}

\newcommand{\vprimal}{\ensuremath{V^{P}}}
\newcommand{\vdual}{\ensuremath{V^{D}}}
\newcommand{\vprimalb}{\ensuremath{V^{P}_B}}
\newcommand{\vdualb}{\ensuremath{V^{D}_B}}

\newcommand{\qual}{\ensuremath{\theta}}
\newcommand{\nqual}{\ensuremath{\state}}
\newcommand{\nquals}{\ensuremath{\states}}

\newcommand{\po}{\ensuremath{p_o}}
\newcommand{\pc}{\ensuremath{p_c}}
\newcommand{\qo}{\ensuremath{q_o}}
\newcommand{\qc}{\ensuremath{q_c}}
\newcommand{\pop}{\ensuremath{\tilde{p}_o}}
\newcommand{\pcp}{\ensuremath{\tilde{p}_c}}
\newcommand{\qop}{\ensuremath{\tilde{q}_o}}
\newcommand{\qcp}{\ensuremath{\tilde{q}_c}}
\newcommand{\pol}{\ensuremath{\overline{p}_o}}
\newcommand{\pcl}{\ensuremath{\overline{p}_c}}
\newcommand{\qol}{\ensuremath{\overline{q}_o}}
\newcommand{\qcl}{\ensuremath{\overline{q}_c}}

\newcommand{\xl}{\ensuremath{\underline{x}}}

\newcommand\beas{\begin{eqnarray*}}
\newcommand\eeas{\end{eqnarray*}}
\newcommand\bit{\begin{itemize}}
\newcommand\eit{\end{itemize}}
\newcommand\ben{\begin{enumerate}}
\newcommand\een{\end{enumerate}}
\newcommand\e{\varepsilon}

\newcommand\ga{\gamma}
\newcommand\ol{\overline}
\newcommand\ul{\underline}
\newcommand{\barray}{\left\{\begin{array}{lll}}
\newcommand{\earray}{\end{array}\right.}
\newcommand{\andif}{&\text{if}&}

\newcommand{\bias}{\ensuremath{b}}
\newcommand{\av}{\ensuremath{\overline{a}}}
\newcommand{\avsq}{\ensuremath{\check{a}}}

\newcommand\tp{\tilde{p}}
\newcommand\tq{\tilde{q}}
\newcommand{\lweight}{\ensuremath{r}}
\newcommand{\Lweight}{\ensuremath{R}}

\newcommand{\tstateni}{\ensuremath{\breve{\omega}_{-i}}}
\newcommand{\tactionni }{\ensuremath{\breve{a}_{-i}}}

\newcommand{\h}{\ensuremath{h}}
\newtheorem{theorem}{Theorem}
\newtheorem{lemma}{Lemma}

\newtheorem{definition}{Definition}

\newtheorem{claim}{Claim}
\newtheorem{prop}{\protect\propositionname}

\newtheorem{assumption}{\protect\assumptionname}
\theoremstyle{plain}

\theoremstyle{plain}
\theoremstyle{plain}
\newtheorem{cor}{\protect\corollaryname}
\theoremstyle{plain}

\providecommand{\assumptionname}{Assumption}
\providecommand{\corollaryname}{Corollary}
\providecommand{\lemmaname}{Lemma}
\providecommand{\propositionname}{Proposition}
\providecommand{\theoremname}{Theorem}

\pgfplotsset{compat=1.17}
\begin{document}
\title{
Information Design in Smooth Games

\footnotetext{Smolin: Toulouse School of Economics, University of Toulouse Capitole and CEPR, \href{mailto:alexey.v.smolin@gmail.com}{\texttt{alexey.v.smolin@gmail.com}}. Yamashita: Osaka University, \href{mailto:tytakuroy@gmail.com}{\texttt{tytakuroy@gmail.com}}. For valuable suggestions and comments, we would like to thank the Coeditor Rakesh Vohra, two anonymous referees, Refine.ink, Dirk Bergemann, Deniz Dizdar, Laura Doval, Philippe Jehiel, Jiangtao Li, Xiao Lin, Elliot Lipnowski, Stephen Morris, Alessandro Pavan, Antonio Penta, Jacopo Perego,  Fedor Sandomirskiy, Ludvig Sinander, Takashi Ui, Xavier Vives, and Alexander Wolitzky, as well as seminar participants at Toulouse School of Economics, Western University, University of Toronto,  University of Surrey, University of Oxford, Hitotsubashi University, Singapore Management University, CMU/Pittsburgh, Pennsylvania State University, MIT/Harvard, Bonn Winter Theory Workshop 2022, SWET 2022, CMid2022, EC 2022, EEA-ESEM 2022,  EARIE 2022, and Venice Winter Theory Workshop 2023. An extended abstract of a previous version of this paper appeared as ``Information Design in Concave Games'' in the proceedings to EC'22. Smolin gratefully acknowledges funding from the French National Research Agency (ANR) under the Investments for the Future program (grant ANR-17-EURE-0010) and through the AI Interdisciplinary Institute ANITI (grant ANR-23-IACL-0002), as well as the hospitality of Northwestern University and Columbia Business School, where parts of this paper were completed. Yamashita gratefully acknowledges funding from the French
National Research Agency (ANR) under the Investments for the Future program (grant
ANR-17-EURE-0010), funding from the European Research Council (ERC) under the European Union's Horizon 2020 research and innovation program (grant 714693), and JSPS KAKENHI (grant 23K01311).}

\author{Alex Smolin, Takuro Yamashita}
}
%
%
\date{April 22, 2026}

\maketitle
\begin{abstract}
We study information design in games where players choose from a continuum of actions and have continuously differentiable payoffs. We show that an information structure is optimal when the equilibrium it induces can also be implemented in a principal-agent contracting problem. Building on this result, we characterize optimal information structures in symmetric linear-quadratic games. With common values, targeted disclosure is robustly optimal across all priors. With interdependent and normally distributed values, linear disclosure is uniquely optimal. We illustrate our findings with applications in venture  capital, Bayesian polarization, and price competition.

\end{abstract}
\thispagestyle{empty}
\setcounter{page}{0}
\vfill
\section{Introduction}
As advances in IT infrastructure and artificial intelligence expand our capacity to collect, process, and generate data, an increasing number of firms, regulators, and individuals must decide which information to supply to strategically interacting agents. Which information should investors receive to allocate capital most efficiently? Which demand signals should be disclosed to improve market outcomes? What are the limits of Bayesian polarization? We show that these and related questions can be tractably addressed within a single framework, and that the optimal policies are simple, intuitive, and robust.

Our main analysis focuses on   concave games of incomplete information, i.e., games in which each player's action lies in a convex set and each player's payoff is weakly concave in that action, but our methods extend to any smooth game. Concave games are a staple of applied economic modeling with fixed information structures, because equilibria can be tractably characterized by first-order conditions. We show that the same considerations render tractable the task of designing the information structure itself.

To solve the information-design problem, we adopt a duality-based approach. The dual can be viewed as an adversarial contracting problem between a principal and an omniscient agent who both observes the state and controls all players' actions. The dual-certification theorem (\autoref{thm:dual_certification}) states that if a state-action distribution emerges in equilibrium under both an information structure (in the information-design problem) and a contract (in the adversarial-contracting problem), then that information structure and that contract are optimal for their respective problems. The contract thus serves as an optimality certificate. Furthermore, it is capable of certifying any optimal information structure (\autoref{prop:universality}), which enables us to establish whether the optimal information structure is unique or, when several are optimal, to identify their common features.

We show that (certifiably) optimal information structures are prior-robust: once optimal under one prior, they remain optimal under any other prior as long as the implemented allocation rule's support is contained within the original support in every state (\autoref{prop:robustness}). This has two consequences. First, if an optimal information structure is fully informative about the state, it remains optimal under all priors. Second, unless the problem is trivial, an optimal information structure cannot induce a full-support action distribution in every state, which argues against adding full-support, extraneous, independent noise to individual signals.

We apply the solution method in two broad, symmetric settings with linear-quadratic payoffs for both the players and the designer. In each case, the certifying contract is symmetric across players and affine in actions, yet it yields qualitatively different optimal information policies.

In the first setting, the state is one-dimensional (\autoref{sec:lqg-one-dim}). We derive the parameters of the certifying contract and use them to establish the optimality of a novel information structure we term \emph{targeted disclosure}: it fully reveals the state to a subset of players while leaving the rest completely uninformed. This structure is remarkably simple and distributionally robust---it remains optimal under any prior. In addition, when the state is normally distributed, \emph{Gaussian coupling}, which adds normal noise to each player's signal that cancels out in aggregate, is optimal as well.

In the second setting, the state is multidimensional with jointly normal components (\autoref{sec:lqg-multi-dim}). We show that a certifying contract can be found within the class of symmetric affine contracts by solving a single-variable minimization problem. This contract certifies  the unique optimality of \emph{linear disclosure}: an information structure that gives each player some linear statistic of the state. The resulting information structure is noise-free and symmetric ex ante.

In \autoref{sec:applications}, we  illustrate these results in three concrete applications. In \autoref{sec:investment}, we analyze optimal capital fundraising when several investors independently choose how much to invest in a project of uncertain quality. We show that the information structure that maximizes the project's expected return is exclusive disclosure, a targeted disclosure in which only one investor is informed. Relative to full or no disclosure, exclusive disclosure prevents the dissipation of returns as the investor pool grows, offering a possible rationale for a common venture-capital practice. 

In \autoref{sec:belief_polarization}, we examine the limits of Bayesian polarization when each player forms a forecast from private information. We show that the information structure that maximizes a natural polarization index is a targeted disclosure that informs exactly half of the players. Hence, maximal polarization occurs when the population splits into two cohorts that are internally uniform but sharply divergent from each other. This finding underscores how media segregation can intensify societal polarization. 

In \autoref{sec:price_competition}, we study optimal price recommendations in a differentiated-product duopoly with linear demand and stochastic demand shocks. We characterize the information structures that maximize any weighted average of consumer and producer surplus. The resulting recommended prices are linear functions of the demand shocks and exhibit discontinuous regime changes with respect to the weight on consumer surplus. This observation underscores the risk that algorithms generating price recommendations---able to shift objectives and adapt policies far faster than human decision-makers---can destabilize markets.

\paragraph{Related Literature}
Our paper contributes to the recent and flourishing literature on information design. Much of  this literature focuses on information design with a single player, called Bayesian persuasion (\cite{rase10}, \cite{kage11}). Popular solution methods are belief based, i.e., they operate within the space of the receiver's belief distributions.\footnote{See, for example, \cite{dwma19}, \cite{diko20}, and \cite{dwko22}. The single-player case also covers scenarios in which there are many players but the information is required to be public.}  A natural continuation of this research agenda is a study of information design in multiplayer games (\cite{bemo16}, \cite{tane19}).
In games, players' beliefs constitute infinite hierarchies, which renders  the belief-based approach less tractable (\citet*{mathevet2020information}). Instead, in games, an action-based approach rooted in the revelation principle is promising, as it frames the design problem as a linear program and enables the use of duality machinery.\footnote{The duality methodology is routinely used in many disciplines to solve optimization problems. In mechanism design, duality methods have been recently used to study optimal delegation (\citet*{amador2006commitment, amador2013theory}), matching (\citet*{chiappori2017partner,galichon2022cupid}),
robust selling mechanisms (\citet{carroll2017robustness, du18, brooks2021optimal}), mediation (\cite{salamanca2021value}, \citet*{ortner2023mediated}), and limited commitment (\citet*{lin2023credible}) among others. For a unified treatment, see \cite{vohra2011mechanism}.}\footnote{Alternatively, one can develop original arguments tailored to the studied problem; see, for example, \cite{arbi19}, \citet*{chan2019pivotal}, \citet*{elliott2022market},  \citet*{arieli2023persuasion}, and \citet*{cast22}.}

\cite{gape18} and \citet*{galperti2023value} employ this action-based approach to study information design in games with finitely many actions, imposing minimal structure on payoffs. They develop an economic interpretation of the Lagrange multipliers associated with Bayes' plausibility as the value of data records and propose the idea of pooling externalities across records; all these observations apply to our setting.

In contrast, we study games with infinitely many actions and impose a concavity structure on payoffs, thus enabling us to rely on a first-order approach for incentives (\cite{holmstrom1979moral}, \cite{mirrlees1999theory}) that leads to more succinct and tractable primal and dual problems. This approach was introduced in a Bayesian persuasion setting by \cite{kolotilin2012optimal, kolotilin2018optimal} and further refined by \citet*{kocowo2022}. These papers study an information-design problem with a single player and a one-dimensional state, identifying when censorship or, respectively, assortative disclosures are optimal. We deepen and extend this approach, adapting it to settings with multiple players and a multidimensional state.

Most of our current understanding of optimal information in   games is drawn from the study of Gaussian signals in games with quadratic payoffs and a normally distributed state (e.g., \cite{anpa07,anpa09}, \cite{bemo13a}, \citet*{bergemann2015information,bergemann2021information}, \cite{ui20lqg}). In that literature, as well as in a vast body of work in macroeconomics and finance, the Gaussian form of players' signals is imposed \emph{ad hoc} for analytical convenience.

Recent arguments  suggest  that Gaussian signals are not only convenient but often optimal among all  information structures. For example, \cite{tamu12,tamu18} established the optimality of a Gaussian signal in a setting with a single player by building on the statistical properties of a covariance matrix of posterior expectations. \citet*{bergemann2017information} and \citet*{miyashita2023lqg} extend this argument to games.\footnote{In their Section 4.4, \citet{bergemann2017information}  argue that Gaussian structures, possibly asymmetric and with extraneous noise, span all implementable covariance matrices of equilibrium actions.} Our results align with these findings and show that in many scenarios the optimal Gaussian structures are symmetric and noise-free; moreover, the certification method allows direct identification of the optimal informational parameters. Furthermore, we provide tools that establish when Gaussian information structures are uniquely optimal.

At the same time, we show that in many settings targeted disclosure is optimal. These information structures are simple to implement and distributionally robust, offering a strong positive message for the information design literature, where existing solutions often lack these properties.
\section{Design Problem}\label{sec:model}
We study a standard  information design problem as presented by \cite{bemo16}, extended to accommodate a continuum of players' actions.
\paragraph{Payoffs}
There are  $N$ players indexed by $i$, $1\leq N<\infty$, and an information designer.  Each player chooses an action $a_{i}\in A_{i}=\reals$.
We denote an action profile by $a\in A=\times_{i}A_{i}$ and
write $(a_{i},a_{-i})$ when highlighting player $i$'s action.

A state $\state$ is distributed over a Polish set $\states$,
according to a full-support prior  $\mu_{0}\in\Delta(\states)$. The action profile and the state jointly determine  payoffs through 
\begin{align}\label{eq:playeri_payoff_function}
u_i:A\times\states\rightarrow\reals,\\
\up:\actions\times\states\rightarrow\reals,\label{eq:designer_payoff_function}
\end{align}
for each player $i$ and for the designer, respectively. The tuple $((A_i,u_i)_{i=1}^N,\mu_0)$ constitutes the \emph{basic game}. 

\paragraph{Information}
The players and the designer start with a commonly known prior belief about the state $\state$ that coincides with the prior $\mu_0$. The designer can provide additional information to players by choosing an information structure $\info=(S,\pi)$
that consists of a measurable signal set $S=\times_{i}S_{i}$  and a likelihood function $\pi\in\Delta(S\times\states)$ that has $\mu_0$ as its state marginal distribution. This information structure determines the sets of private signals the players can observe and, through the likelihood function, their informational content. 

First, the designer chooses an information structure $\info$. Second, the state $\state$ and the signal profile $s$ are realized according to $\info$. Finally, each player privately observes his  signal $s_{i}$ and chooses an action $a_{i}$.

We will regularly refer to two benchmark information structures: \emph{full disclosure}, with $S_i = \Omega$ and $s_i \equiv \omega$ for all $i$; and \emph{no disclosure}, with $|S_i| = 1$ for all $i$.

\paragraph{Equilibrium}
The basic game together with the information structure chosen by the designer determine a Bayesian game of incomplete information. In that game, each player's behavior is described by a strategy that maps any received signal to a possibly random action, $\sigma_i:S_i\rightarrow \Delta(A_i)$, and we consider as an equilibrium concept a Bayes Nash equilibrium:
\begin{definition}
\label{def:equilibrium}\textup{(Bayes Nash Equilibrium)} For a given information structure $\info$,  a strategy profile $\sigma=(\sigma_1,\dots,\sigma_N)$ constitutes a  Bayes Nash  equilibrium if 
\begin{align}\label{eq:BNE_conditions}
\expect_{\info,\sigma_i,\sigma_{-i}}[u_i(a_i,a_{-i},\state)]\geq\expect_{\info,\sigma_i',\sigma_{-i}}[u_i(a_i',a_{-i},\state)]
\end{align}
for all i and $\sigma_i': S_i\rightarrow \Delta(A_i)$.\footnote{We write $\expect_{\info,\sigma_i,\sigma_{-i}}[\cdot]$ for the expectation under information structure $\info$ and   strategy profile $(\sigma_i,\sigma_{-i})$. Throughout the paper an integral is left undefined whenever the integrand is not integrable with respect to the relevant measure.
} 
\end{definition}

An information structure and a strategy profile determine  a distribution over the action profiles in each state $\alpha:\states\rightarrow\Delta(A)$, which we call an \emph{allocation rule}, and the corresponding designer's expected payoff $\expect_{\info,\sigma}[\up(\action,\state)]$. The \emph{value} of an information structure is defined as the maximal designer's expected payoff that can arise in equilibrium of the induced  game: if the  game has multiple equilibria, the designer can choose the one she prefers, whereas if no equilibrium exists, the value is undefined. An information-design problem consists of finding an information structure with a maximal value without placing any additional restrictions on the sets of signals or the likelihood function, apart from  a mild ``admissibility'' condition. (This condition, and other omitted formal details, are deferred to \autoref{app:A}.)
\begin{definition}\textup{(Optimal Information Structure)}
An information structure is optimal if there does not exist an information structure with a strictly higher value. 
\end{definition}

The search for an optimal information structure is complicated by the scale of the basic game: multiple players, after receiving private signals, choose actions from a continuum while anticipating one another's behavior. Our central simplifying assumption is:
\begin{assumption}
\label{ass:concavity}\textup{(Concave Payoffs)} For all $i=1,\dots,N$,
$\state\in\states$, and $a_{-i}\in A_{-i}$, $u_{i}(a_{i},a_{-i},\state)$
is continuously differentiable in  $a_{i}$, weakly concave in $a_{i}$, and
obtains its maximum at some finite value. 
\end{assumption}

We call a basic game in which \autoref{ass:concavity} is satisfied a \emph{concave game}.\footnote{This notion of a concave game is related to but distinct from the notion of a concave game of \cite{rose65}. 
In particular, it requires neither differentiability nor continuity of a player's payoff in the other players' actions or the state.} In a concave game, for any player $i$ and equilibrium belief $\belief\in\Delta(A_{-i}\times\states)$ about the others' actions and the state, a best response $a_{i}^{*}$ exists and satisfies the first-order condition:
\begin{gather}\label{eq:foc_belief}
\left.\frac{\partial}{\partial a_i}\expect_{\belief}\left[u_{i}(a_{i},a_{-i},\state)\right]\right|_{a_i=a_i^{*}}=\expect_{\belief}\left[\left.\frac{\partial}{\partial a_i}u_{i}(a_{i},a_{-i},\state)\right]\right|_{a_i=a_i^{*}}\triangleq\mathbb{E}_{\belief}\left[\dot{u}_{i}(a_{i}^{*},a_{-i},\state)\right]=0,
\end{gather}
where we denote the player's marginal payoff function by
\begin{align*}
\dot{u}_{i}(a,\state)\triangleq\frac{\partial u_{i}(a,\state)}{\partial a_i}.
\end{align*}
An equilibrium under any given information structure is characterized by a system of such conditions, one for each player's signal.

The information-design problem can be simplified by appealing to the revelation principle (\cite{myer82}): it is without loss of generality to focus on \emph{direct information structures}  that inform each player about a recommended action $S=A$ and are such that all players are obedient, i.e., are willing to follow the recommendations. Each direct information structure corresponds to a measure $\pi\in\Delta(A\times\states)$, and the information-design problem can be  formulated as a constrained maximization over these measures:
\begin{alignat}{1}
\vprimal\triangleq\sup_{\pi\in\Delta(A\times\states)} & \int_{A\times\states}v(a,\state)\textrm{d}\pi\label{problem:sender_primal}\\
\textrm{s.t.\ } & \int_{A'_{i}\times A_{-i}\times\states}\dot{u}_{i}(a,\state)\textrm{d}\pi=0\quad\forall\,i=1,\dots,N,\textrm{measurable\ }A'_{i}\subseteq A_{i},\label{eq:obedience_constraints}\\
 & \int_{A\times\states'}\textrm{d}\pi=\int_{\states'}\textrm{d}\mu_{0}\quad\forall\,\textrm{measurable\ } \states'\subseteq\states.\label{eq:feasibility_constraints}
\end{alignat}
Constraints (\ref{eq:obedience_constraints}) are a proper formulation of first-order conditions (\ref{eq:foc_belief}) in light of a continuum of recommended actions. These constraints capture players' obedience  and effectively
require that for each player $i$, the marginal of $\pi$
on $A_{i}$ weighted by the marginal utilities equals zero measure.  Constraints (\ref{eq:feasibility_constraints}) capture Bayes' plausibility and,
likewise, require that the marginal of $\pi$ on $\states$ equals
the prior $\mu_{0}$.
\section{Solution Method}\label{sec:general_analysis}

Problem (\ref{problem:sender_primal}--\ref{eq:feasibility_constraints}) is linear in $\pi$, yet unwieldy to solve directly. In the spirit of linear programming, we view it as a \emph{primal problem} and call any $\pi\in\Delta(A\times\states)$ a \emph{primal measure}.  If a primal measure satisfies the constraints of the primal problem, then we call that measure \emph{implementable by information}. We can construct a  dual problem  as follows (e.g., \cite{anderson1987linear}):
\begin{alignat}{1}
\vdual\triangleq\inf_{\lambda\in\times_{i}\meas(A_{i}),\gamma\in\meas(\states)} & \ \int_{\states}\gamma(\state)\textrm{d}\mu_{0}\label{problem:sender_dual}\\
\textrm{s.t.}\ \sum_{i=1}^{N} & \lambda_{i}(a_{i})\dot{u}_{i}(a,\state)+\gamma(\state)\geq v(a,\state)\ \forall\,a\in A,\state\in\states,\nonumber 
\end{alignat}
where $\meas(X)$ denotes the space of measurable real-valued functions on $X$. The minimization arguments, the dual variables $(\lambda,\gamma)$,  represent the Lagrange multipliers associated with  the primal incentive constraints (\ref{eq:obedience_constraints}) and the feasibility constraints (\ref{eq:feasibility_constraints}), respectively.

The problem (\ref{problem:sender_dual}) is a generalization of the dual problem of \cite{kolotilin2018optimal} and \cite{kocowo2022} to multiple players. By the arguments analogous to those of \cite{galperti2023value}, the optimal $\gamma^*(\state)$ measures the marginal benefit of increasing the frequency of state $\state$. Importantly, the optimal $\gamma^*(\state)$ can be solved away: the objective in (\ref{problem:sender_dual}) is additively separable in $\gamma(\state)$, and the constraints at different states $\state$ are linked only through the variables $\lambda$. Hence, for any $\lambda$ and $\state$, the optimal choice of $\gamma(\state)$ is the smallest value consistent with the dual constraints:
\begin{align*}
\gamma^{*}(\state;\lambda)=\sup_{a\in A}\ud(a,\state,\lambda),
\end{align*}
where  the \emph{dual payoff function} $\ud$ is defined as
\begin{align}\label{eq:dual_payoff}
   \ud(a,\state,\lambda)\triangleq v(a,\state)-\sum_{i=1}^{N}\lambda_{i}(a_{i})\dot{u}_{i}(a,\state).
\end{align}
As a result, the problem (\ref{problem:sender_dual}) can be restated as follows:
\begin{align}\label{problem:adversarial_contracting}
\vdual=\inf_{\lambda\in\times_{i}\meas(A_{i})} \expect[\sup_{a\in A}\ud(a,\state,\lambda)].
\end{align}

Problem (\ref{problem:adversarial_contracting}) admits a simple economic interpretation as a \emph{contracting problem} between a \emph{dual principal} and a   \emph{dual agent}.  First, the principal chooses an incentive contract $\lambda$ that consists of $N$ functions $\lambda_i(a_i)$ and determines the agent's payoff according to (\ref{eq:dual_payoff}). 
Second, the state $\state$ is realized. Finally, the agent perfectly observes the state and chooses the whole action profile $a\in A$ to maximize his payoff. The contracting is \emph{adversarial} in that the principal aims to minimize the  agent's expected payoff. 
 
If the best responses exist at all states and induce the joint action-state measure $\pi(a,\state)$, then we say that $\lambda$ \emph{implements  $\pi$ by incentives} and that \emph{$\pi$ is implementable by incentives, by contract $\lambda$}. 

\begin{theorem}
\textup{(Weak Duality. Dual Certification)} \label{thm:dual_certification} If $\pi\in\Delta(A\times\states)$ is implementable by information and is implementable by incentives by contract $\lambda$, then (i) $\pi$ solves the information-design problem, (ii) $\lambda$ solves the adversarial-contracting problem, and (iii) $\vprimal=\vdual$.
\end{theorem}

\autoref{thm:dual_certification} offers a solution method based on optimality certification. 
When the conditions of the theorem hold,  we  say that $\lambda$ is a (dual) \emph{certificate} of $\pi$, that $\lambda$  \emph{certifies} the optimality of $\pi$, and that $\pi$ is  \emph{certifiably optimal}. Similarly, if an information structure induces a certifiably optimal measure, then  we call that information structure  \emph{certifiably optimal}. 

We highlight two general properties of certificates. First, every certificate $\lambda^*$ has a clear economic meaning: it represents Lagrange multipliers associated with the obedience constraints (\ref{eq:obedience_constraints}). Accordingly, $\lambda^*_{i}(a_{i})$ measures the marginal value for the information designer from perturbing the obedience constraint of action $a_{i}$.
Second, each certificate is ``universal'':
\begin{prop}\label{prop:universality}\textup{(Universality)}
If $\lambda$ certifies the optimality of measure $\pi$, and $\pi'$ is another optimal measure, then $\lambda$ also certifies the optimality of measure $\pi'$.
\end{prop}

By \autoref{prop:universality}, a certificate does more than certify a single information structure---it constrains the entire set of optima. Every optimal measure must be certified by  the same certificate; equivalently, it must be the dual agent's best response to the same contract in the dual problem. We exploit this property  in  \autoref{sec:lqg-one-dim}, where multiple optima exist, to isolate their common features, and in  \autoref{sec:lqg-multi-dim} to show that the optimal measure is unique.

The difference $\vdual-\vprimal$ between the optimal values of  primal and  dual problems is nonnegative and constitutes a \emph{duality gap}. The solution to the information-design problem can be certified if and only if (i)  solutions to both primal and dual problems exist and (ii) the duality gap is equal to zero, $\vdual=\vprimal$. Thus, either all optimal information structures can be certified or none of them can.

Any constructed certificate, by its very existence, proves that the duality gap is zero and that both primal and dual solutions exist. However, it may be useful to know \emph{a priori} whether one can expect to find such a certificate in a given problem.  
\begin{claim}
\textup{(Strong Duality)} \label{thm:strong_duality} If each $A_i$ and $\Omega$ are compact, and $v$ and each $\dot{u}_i$ are continuous on $\Omega\times A$, then $\vdual=\vprimal$.
\end{claim}
\autoref{thm:strong_duality} follows from standard Fenchel-Rockafellar arguments. We record the result mainly as a benchmark, suggesting that strong duality is to be expected in regular environments. Many applications below feature noncompact action spaces. In those environments, a general strong-duality theorem would require additional coercivity and integrability assumptions to rule out mass escaping to infinity and to ensure that the dual certificates remain well behaved. Rather than impose such conditions abstractly, in the applications below we verify strong duality constructively by exhibiting explicit certificates.

\section{Robustness of Optimal Information Structures}\label{sec:on-certifiable}
By \autoref{thm:dual_certification}, an allocation rule induced by a certifiably optimal information structure must be chosen freely by a fully informed dual agent. This observation has two consequences. First, the prior is irrelevant for the implementability of an allocation rule by incentives since the prescribed action profiles must be optimal state-by-state. Second, if the optimal allocation rule randomizes over several action profiles at a given state, the dual agent must be indifferent among those profiles and could therefore randomize over them with any probabilities. Hence, only the support of the action profiles matters for implementability by incentives:

\begin{prop}\label{prop:robustness}
\textup{(Robustness)}
Let two concave
information-design problems  differ  only in their priors, if at all. Let $\pi_1\in\Delta(\actions\times\states)$ be certifiably optimal in the first problem. If $\pi_2\in\Delta(\actions\times\states)$ is implementable by information in the second problem and $\textup{supp}\,\pi_2(\cdot\mid\state)\subseteq\textup{supp}\,\pi_1(\cdot\mid\state)$  for all $\state\in\states$, then $\pi_2$ is certifiably optimal in the second problem.
\end{prop}

\autoref{prop:robustness} shows that certifiably optimal information structures are, to some extent, \emph{prior-robust}: once optimal under one prior, an information structure remains optimal under any other prior, provided it still implements an allocation rule whose support is no larger in every state.
 It is specific to our setting, which features a continuum of actions and thus more flexible players' best responses, and does not hold in generic games with finitely many actions (cf. \citet{kage11}).

 Generally, the larger the support is, the easier it is to construct multiple information structures that implement allocation rules within that support. In the extreme case, if the action support covers the whole action space, then the support condition of \autoref{prop:robustness} has no bite, and any information structure can be certified to be optimal.
\begin{cor}
\textup{\label{cor:ind_noise}(Full-Support Noise)} If measure $\pi\in\Delta(\actions\times\states)$ is  certifiably optimal and  $\textup{supp}\,\pi(\cdot\mid\state)=A$ for all $\state\in\states$, then any information structure is certifiably optimal.
\end{cor}
\autoref{cor:ind_noise} presents a case against using extraneous noises that induce full-support action profiles in concave games with finitely many players.
The  information structures that employ such noises can never be certifiably optimal, except in  trivial cases in which the designer's expected payoff is invariant to the information provided. This finding resonates with the analysis of \cite{tane19}, who studied a two-player binary setting and showed that sending conditionally independent signals is never strictly optimal, as well as with the analysis of \cite{cast22}, who established optimality of partitional signals in a class of games.  However, such extraneous independent noises may optimally appear in the limit information structure as the number of players grows to infinity, as we discuss in \autoref{sec:extensions}.

Furthermore, \autoref{prop:robustness} enables us to assess the optimality of full state transparency. We call an information structure \emph{fully informative about the state} if each player can deduce the state with certainty from her private signal. There could be many such structures, each differing in the state-by-state coordination of players' actions, including full disclosure. However, any such structure can implement the same allocation rule under all priors.
\begin{cor}
\textup{\label{cor:full-and-no-disclosure}(Full State Information)} If a certifiably optimal information structure is  fully informative about the state, then it is certifiably optimal under all priors.
\end{cor}
\autoref{cor:full-and-no-disclosure} shows that, once full disclosure is optimal, it is extremely robust: it remains optimal under any prior (cf. \cite{jehiel2015transparency}).

\section{Linear-Quadratic Games: Common Value}\label{sec:lqg-one-dim}
\renewcommand*{\appendixautorefname}{Online Appendix} 
The certification approach developed in the previous section applies to any concave game (see Online Appendix) and, more generally, to any smooth game (\autoref{sec:extensions}).
 In this section, we demonstrate the approach in symmetric linear-quadratic games with a common state. We first develop general results and then illustrate them through two applications: capital fundraising and expectation polarization.

\renewcommand*{\appendixautorefname}{Appendix} 

We define a \emph{linear quadratic symmetric game with a common state} as a game with \(\Omega \subseteq \mathbb{R}\), and the following payoff structure:
\begin{align}
    u_i(a,\omega)&=(\omega+\bias_1) a_i+\frac{1}{2} (q^o-\frac{q^c}{N})a_i^2+q^c a_i \av+f_1(a_{-i},\omega),\label{eq:common-payoff-gen-1}\\
    v(a,\omega)&=(\h\omega+\bias_2) \av+p^o \avsq+p^c \av^2+f_2(\omega),\label{eq:common-payoff-gen-2}
\end{align}
where $\av\triangleq \sum_{j}a_j/{N}$, $\avsq\triangleq \sum_{j}a_j^2/{N}$, $f_1$ and $f_2$ are arbitrary functions of the indicated arguments, $\h\geq 0$, and $q^o$, $q^c$, $p^o$, $p^c$ satisfy the  concavity conditions: 
\begin{gather}
    q^o<0,\ q^o+q^c<0,\label{eq:common_concavity_cond_1}\\
    q^o p^c > p^o q^c.\label{eq:common_genericity_cond}
\end{gather}
Condition (\ref{eq:common_concavity_cond_1}) ensures that $u_i$ is strictly concave in $a_i$ for all $N\geq1$, whereas condition (\ref{eq:common_genericity_cond}) ensures concavity of the dual payoff under conjectured certificates.
This game is strategically equivalent to, and thus admits the same solution as, a simpler  \emph{normalized} game, in which $\expect[\omega]=0$,
\begin{align*}
    u_i(a,\omega)&=\omega a_i+\frac{1}{2} (q^o-\frac{q^c}{N})a_i^2+q^c a_i \av,\\
    v(a,\omega)&=(\h\omega+\bias) \av+p^o \avsq+p^c \av^2,
\end{align*}
where $\h$, $q^o$, $q^c$, $p^o$, $p^c$ are as in the original game and $b=\h\expect[\omega]+\bias_2-2(\expect[\omega]+\bias_1)(p^o+p^c)/(q^o+q^c)$. Therefore, we can focus on an analysis of the normalized game.

To find optimality certificates, observe that the players' marginal payoffs are linear:
\begin{align}\label{eq:FOC_common}
   \dot{u}_i(a,\omega)=\omega+q^o a_i+q^c\av,
\end{align}
and the dual payoff function takes the form 
\begin{align}\label{eq:lqg-dual-payoff}
\ud(a,\state,\lambda)=(\h\omega+\bias) \av+p^o \avsq +p^c \av^2 -\sum_{i=1}^N \lambda_i(a_i)(\omega+q^o a_i+q^c\av).
\end{align}

Given the linear-quadratic nature of the environment, we conjecture that an optimal allocation rule is linear in the state. Because the allocation rule must be a best response given the dual payoff (\ref{eq:lqg-dual-payoff}), it is natural to posit a certificate that is affine and symmetric across players:
\begin{align}\label{eq:affine-contract}
    \lambda_i(a_i)=-\frac{1}N(x a_i+x_0).
\end{align}
With this certificate, the dual payoff function becomes
\begin{align*}
    \ud(a,\state,x,x_0)=x_0\omega+((\h+x)\omega+\bias+x_0(q^o+q^c)) \av+(p^c+x q^c) \av^2+(p^o+q^o x) \avsq .
\end{align*}
If the coefficient in front of \( \avsq \) is nonzero, \( p^o +  q^o x\neq 0 \), then the dual best-response is uniquely defined, linear in state, and symmetric across players; thus, such \( x \) could certify only the optimality of full disclosure or no disclosure. To certify the optimality of partial disclosure, we must have 
\begin{gather*}
x^*=-\frac{p^o}{q^o},\\
\ud(a,\state,x^*,x_0)=x_0\omega+\left(\left(\h-\frac{p^o}{q^o}\right)\omega +\bias+x_0(q^o+q^c)\right)\av+\left(p^c-\frac{p^o q^c}{q^o}\right) \av^2.
\end{gather*}
With this choice of $x$, the dual payoff function is a function solely of  the  average action $\av$. Because of (\ref{eq:FOC_common}) and $\expect[\omega]=0$, any incentive compatible expected average action must equal to $0$; therefore  we must have  
\begin{gather}
x_0^*=-\frac{\bias}{q^o+q^c},\notag \\
\av^*(\omega)=\frac{p^o-\h q^o}{2(p^c q^o-p^o q^c)}\omega.\label{eq:common-optimal-aggregate}
\end{gather}
Any allocation rule that results in this average action  is  implementable by incentives by the affine contract (\ref{eq:affine-contract}). By \autoref{thm:dual_certification}, if the same allocation rule can be implemented by information, then that allocation rule is optimal.\footnote{This observation resonates with the research on robust information design in regime change games, such as \citet*{inpa23} and \citet*{morris2023implementation}, which is fundamentally concerned with multiple informational implementations of the same aggregate behavior.} A natural way to implement such a linear average action is to fully inform a subset of players: 
\begin{definition}{\textup{(Targeted Disclosure)}}
For $k\in\{0,1,\dots,N\}$, a \emph{$k$-targeted disclosure} is an information structure that fully reveals the state to $k$ players while providing no information to others. 
\end{definition}
A $0$-targeted disclosure is  no disclosure and $N$-targeted disclosure is full disclosure. 
Under a $k$-targeted disclosure, there exists an equilibrium where each uninformed player plays $a_i^{ND}\equiv 0$, whereas each informed player plays
\begin{align*}
a_i^I(\omega)=-\frac{N}{q^c k+q^o N}\omega,
\end{align*}
resulting in the average action
\begin{align}\label{eq:common-k-aggregate}
\av(\omega)=\frac{k}{N}a_i^I(\omega)=-\frac{k}{q^c k+q^o N}\omega.
\end{align}
\begin{theorem}{\textup{(Optimality of Targeted Disclosure)}}\label{thm:lqg-common-targeted}
In a normalized linear-quadratic symmetric game with a common state, if 
\begin{align}\label{eq:k-opt}
k^*\triangleq\frac{q^o(\h q^o-p^o)}{q^o(2p^c-\h q^c)-p^oq^c}N
\end{align}
is in $\{0,1,\dots,N\}$, then $k^*$-targeted disclosure is optimal. If $k^*\notin[0,N]$,\footnote{This includes the case when the denominator in (\ref{eq:k-opt}) is zero.} then  either (i) $p^o<\h q^o$, in which case no disclosure is optimal, or (ii)  $(\h q^o-p^o)(q^o+q^c)>2(p^c q^o-p^o q^c)$, in which case full disclosure is optimal.
\end{theorem}
\begin{proof}
If $k$  is in $\{0,1,\dots,N\}$, then by \autoref{thm:dual_certification}, comparing (\ref{eq:common-optimal-aggregate}) and (\ref{eq:common-k-aggregate}), a $k$-targeted disclosure is optimal if 
\begin{align*}
&-\frac{k}{q^c k+q^o N}=
\frac{p^o-\h q^o}{2(p^c q^o-p^o q^c)}\\
&\Leftrightarrow
k=\frac{q^o(\h q^o-p^o)}{q^o(2p^c-\h q^c)-p^o q^c}N.
\end{align*}

If $k\notin[0,N]$, then given the  conditions (\ref{eq:common_concavity_cond_1})
, either (i) or (ii)  holds.\footnote{Note that conditions (i) and (ii) are not equivalent to $k^*<0$ and $k^*>N$. Examples exist in which full disclosure is optimal even when $k^* < 0$, and others in which no disclosure is optimal even when $k^* > N$.} In case of (i), the optimality of no disclosure can be certified by an affine contract with $(x,x_0)=(-\h,-\bias/(q^o+q^c))$. In case of (ii), the optimality of full disclosure can be certified by an affine contract with $(x,x_0)=(\h- 2(p^o+p^c)/(q^o+q^c),-\bias/(q^o+q^c))$. (See Appendix \ref{app:section4_targeted} for details.)
\end{proof}

\autoref{thm:lqg-common-targeted} highlights the importance of targeted disclosure in linear-quadratic games with a common state. First, if $k^*\notin(0,N)$, then either $0$-targeted disclosure or $N$-targeted disclosure is optimal. Second, if $k^*$ is in $\{0,1,\dots,N\}$, then $k^*$-targeted disclosure is exactly optimal, and the parameter space that yields it spans a wide range of economically relevant settings. 
Third, if $k^*\in(0,N)$ but is not an integer, a $\lceil k^* \rceil$-targeted disclosure is asymptotically optimal. Indeed, the affine contract above provides, by weak duality, an upper bound on the designer's value, and under any obedient allocation the contract term has expectation zero. Under this contract, the dual payoff depends only on the average action and is maximized at (\ref{eq:common-optimal-aggregate}). Because the average action induced by $k$-targeted disclosure depends continuously on $k/N$, and $\lceil k^*\rceil/N \to k^*/N$, the payoff under $\lceil k^*\rceil$-targeted disclosure converges to the optimum as $N$ grows.

The optimality of targeted disclosure is notable for three reasons. First, when $k^{*}\in[1,N-1]$, the policy is asymmetric even though the underlying environment is symmetric. Second, because the argument never relies on the prior, the same targeted-disclosure policy is optimal for \emph{all} priors (cf. \autoref{prop:robustness}). Third, the policy is remarkably simple to implement. The latter two properties highlight the exceptional practical applicability of targeted disclosure.

At the same time, optimal information structures need not be unique. When targeted disclosure is optimal, the designer is free to choose which players receive information. That choice leaves aggregate equilibrium outcomes unchanged but shifts surplus among players.
For example, an ex-ante symmetric variant of $k^{*}$-targeted disclosure first publicly draws a random subset of $k^{*}$ players and then applies $k^{*}$-targeted disclosure to that subset.
Furthermore, qualitatively distinct information structures can also be optimal, specifically when the state is normally distributed.

For the rest of this section, assume $N\geq 2$.
 
\begin{definition}{\textup{(Gaussian Coupling)}}
For $\beta\in\reals$ and $\sigma^2>0$, a \emph{$(\beta,\sigma^2)-$Gaussian coupling} is an information structure such that for all $i$ and $\omega$,
\begin{align*}
    s_i=\beta \omega +\e_i-\frac{1}{N-1}\sum_{j\neq i}\varepsilon_j,
\end{align*}
where each noise term $\e_i\sim N(0,\sigma^2)$ is independent from $\omega$ and $\e_j$ for $j\neq i$.
\end{definition}

A Gaussian coupling distorts the state by adding Gaussian noise to each player's signal, with the individual noises coupled in such a way that they vanish upon aggregation---so that the sum of all signals is a deterministic function of the state $\omega$. When $\beta \neq 0$, knowledge of all signals perfectly reveals the state.
However, observing only $s_i$ provides imperfect information about the state; in this sense, a Gaussian coupling  splits the complete state information across players. When the state is normally distributed, a Gaussian coupling corresponds to a Gaussian information structure---that is, the signals and the state are jointly normally distributed.

\begin{theorem}{\textup{(Optimality of Gaussian Coupling)}}\label{thm:lqg-common-gaussian}
In a normalized linear-quadratic symmetric game with a common state, if $\omega\sim N(0,\sigma_\omega^2)$ and $k^*$ as defined in (\ref{eq:k-opt}) lies in $(0,N)$,\footnote{By \autoref{thm:lqg-common-targeted}, if $k^*\notin(0,N)$, then either full disclosure or no disclosure is optimal.} then a $(\beta,\sigma^2)-$Gaussian coupling  is optimal, where
\begin{align}
\beta&=\frac{p^o-\h q^o}{2(p^c q^o-p^o q^c)},\label{eq:gaussian-coupling-beta}\\
\sigma^2&=\frac{N-1}{N}\frac{\beta(1+(q^c+q^o)\beta)}{-q^o}\sigma_\omega^2\label{eq:gaussian-coupling-sigma}.
\end{align}
In the optimal equilibrium, $a_i\equiv s_i$.
\end{theorem}
\begin{proof}
Consider $(\beta,\sigma^2)-$Gaussian coupling with $(\beta,\sigma^2)$ as defined by (\ref{eq:gaussian-coupling-beta}) and (\ref{eq:gaussian-coupling-sigma}). If each player plays $a_i\equiv s_i$, then $\av(\omega)=\beta\omega=\av^*(\omega)$, as defined in (\ref{eq:common-optimal-aggregate}). By \autoref{thm:dual_certification}, if $a_i\equiv s_i$ is incentive-compatible and $\sigma^2>0$, then the $(\beta,\sigma^2)-$Gaussian coupling is optimal.

To show incentive compatibility, observe that $\omega+q^o a_i+q^c\av$ and $a_i$ are jointly normally distributed. Therefore, the incentive compatibility is equivalent to 
\begin{align*}
    \expect[(\omega+q^o a_i+q^c\av)a_i]&=0,&&\\
    \expect[(\omega+q^o a_i+q^c \beta \omega)a_i]&=0,&&\\
    (1+q^c\beta)\beta \sigma_\omega^2+q^o\left(\beta^2\sigma_\omega^2+\sigma^2+\frac{1}{N-1}\sigma^2\right)&=0,&&
\end{align*}
which, after rearrangement, is equivalent to (\ref{eq:gaussian-coupling-sigma}).

Finally, it is straightforward to verify that if $k^*\in(0,N)$, then $\sigma^2>0$; see Appendix \ref{thm:lqg-common-gaussian-app} for details.
\end{proof}

\autoref{thm:lqg-common-gaussian} highlights an alternative way of providing information optimally---by carefully coupling the players' individual noises. In the case of a normally distributed state, this information structure comes closest to the existing Gaussian information design literature (e.g., \cite{anpa07,anpa09}, \cite{bemo13a}, \citet*{bergemann2015information,bergemann2021information}, \cite{ui20lqg}). However, the optimality of a Gaussian coupling depends crucially on the prior's Gaussian form, making this solution less robust than targeted disclosure. Accordingly, in the applications that are presented in \autoref{sec:investment} and \autoref{sec:belief_polarization}, we focus on optimal targeted disclosure. Nonetheless, we encourage the reader to bear in mind the fact that if the state is normally distributed in those applications, then providing information via a Gaussian coupling is also optimal.

\section{Linear-Quadratic Games: Interdependent Values}\label{sec:lqg-multi-dim}
In this section, we apply the certification approach to strategic environments with interdependent values and thus with a multidimensional state. For tractability, we focus on jointly normally distributed state components. As in the previous section, we first present general results and then illustrate them with an economic application: informing competitive pricing.

Formally, we study symmetric games with  $\omega\sim N(\mu \textbf{1},\sigma^2 M(1,\rho))$, where $M(x,y)$ denotes the $N\times N$ matrix whose diagonal entries equal $x$ and off-diagonal entries equal $y$, and $\rho\in(-1/(N-1),1)$. The players' payoffs are
\begin{align}\label{eq:payoffs-general-many-1}
    u_i(a,\omega)&=(\omega_i p^o+\tstateni p^c+\bias_1)a_i + \frac{q^o}{2}a_i^2+ q^ca_i\tactionni +f_1(a_{-i},\omega),\\
    v(a,\omega)&=\sum_{i=1}^N(\omega_i \tp^o+\bias_2)a_i+(\sum_{i=1}^N\sum_{j\neq i}\omega_ia_j)\tp^c+(\sum_{i=1}^N a_i^2)\tq^o+(\sum_{i=1}^N\sum_{j\neq i}a_ia_j)\tq^c+f_2(\omega),\label{eq:payoffs-general-many-2}
\end{align}
where $\tstateni\triangleq\sum_{j\neq i}\omega_j$, $\tactionni \triangleq\sum_{j\neq i}a_j$, $f_1$ and $f_2$ are arbitrary functions of the indicated arguments, and $q^o$, $q^c$, $p^o$, $p^c$, $\tq^o$, $\tq^c$, $\tp^o$, $\tp^c$ are commonly known parameters that satisfy the  concavity and ``genericity'' conditions:
\begin{align}
     q^o-q^c<0,\ q^o+(&N-1)q^c<0,\    p^o>|p^c|\label{eq:many_concavity_cond_1},\\
     q^o\tq^c&\neq q^c\tq^o,\label{eq:many_gener_1}\\
(q^o-q^c)(\tp^o-\tp^c)&\neq (\tq^o-\tq^c)(p^o-p^c),\label{eq:many_gener_2}\\
(q^o+(N-1)q^c)(\tp^o+(N-1)\tp^c)&\neq (\tq^o+(N-1)\tq^c)(p^o+(N-1)p^c)\label{eq:many_gener_3}.
\end{align}
Condition (\ref{eq:many_concavity_cond_1}) ensures that $u_i$ is strictly concave in $a_i$ for all $N\geq1$, whereas other conditions avoid knife-edge cases (e.g., division by zero).

This game is strategically equivalent  to---and thus admits the same solution as---a simpler  \emph{normalized} game, in which $\omega\sim N(0,M(1,\rho))$,
\begin{align*}
    u_i(a,\omega)&=(\omega_i p^o+\tstateni p^c)a_i + \frac{q^o}{2}a_i^2+ q^ca_i\tactionni ,\\
    v(a,\omega)&=\sum_{i=1}^N(\omega_i \tp^o+\bias)a_i+(\sum_{i=1}^N\sum_{j\neq i}\omega_ia_j)\tp^c+(\sum_{i=1}^N a_i^2)\tq^o+(\sum_{i=1}^N\sum_{j\neq i}a_ia_j)\tq^c,
\end{align*}
and where $q^o$, $q^c$, $p^o$, $p^c$, $\tq^o$, $\tq^c$, $\tp^o$, $\tp^c$ are the same as in the original game.  Therefore, we can focus on the analysis of the normalized game.
Given the greater complexity of the present setting, we proceed slightly differently from the previous section. We still rely on affine contracts for certification; however, instead of calculating the optimal contract parameters in closed form, we invoke an envelope theorem. We show that an optimal symmetric affine contract implements an obedient allocation rule, thereby certifying both its global optimality in the dual problem and the optimality of the corresponding measure in the primal problem.

Formally, observe that the players' marginal payoffs are linear:
\begin{align}\label{eq:FOC_many}
   \dot{u}_i(a,\omega)=\omega_i p^o+\tstateni   p^c+q^oa_i+q^c\tactionni ,
\end{align}
and the dual payoff function is
\begin{align}\label{eq:lqg-dual-payoff-2}
\ud(a,\state,\lambda)=v(a,\state)-\sum_{i=1}^N \lambda_i(a_i)(\omega_i p^o+\tstateni p^c+q^oa_i+q^c\tactionni ).
\end{align}

Given the linear-quadratic nature of the environment, we conjecture that an optimal allocation rule is linear in the state. As this allocation rule must be a best-response given the dual payoff (\ref{eq:lqg-dual-payoff-2}), we posit a certificate that is affine and symmetric across players:
\begin{align}
    \lambda_i(a_i)=-(x a_i+x_0).
\end{align}
With this certificate, the dual payoff function can be written as 
\begin{align*}
    \ud(a,\state,x,x_0)&\sim\omega^T M(\tp^o+x p^o,\tp^c+x p^c)a+a^TM(\tq^o+x q^o,\tq^c+x q^c)a\\
&+(\bias+x_0(q^o+(N-1)q^c))\atotal,
\end{align*}
where we omitted an action-independent term $x_0(p^o+(N-1)p^c)\sum_{i=1}^N \omega_i$. Because of (\ref{eq:FOC_many}) and $\expect[\omega]=0$, any incentive-compatible expected individual action must equal zero, and therefore we conjecture that the coefficient in front of $\atotal$ equals zero and thus 
\begin{align}
x_0^*&=-\frac{\bias}{q^o+(N-1)q^c},\notag \\
 \ud(a,\state,x,x_0^*)&\sim\omega^T M(\tp^o+x p^o,\tp^c+x p^c)a+a^TM(\tq^o+x q^o,\tq^c+x q^c)a\notag\\
 &\sim\omega^T M(\ol{p}^o(x),\ol{p}^c(x))a+a^TM(\ol{q}^o(x),\ol{q}^c(x))a.\label{eq:many-dual-payoff}
\end{align}
where we defined 
\begin{align}
    \ol{q}^o(x)&\triangleq\tq^o+x q^o,\ \ol{q}^c(x)\triangleq\tq^c+x q^c,\\ \ol{p}^o(x)&\triangleq\tp^o+x p^o,\ \ol{p}^c(x)\triangleq\tp^c+x p^c.
\end{align}

For the optimal allocation rule $a^*(\omega)$ to maximize (\ref{eq:many-dual-payoff}), $M(\ol{q}^o(x),\ol{q}^c(x))$ must be negative semi-definite; otherwise, the value of the dual payoff diverges to $+\infty$.
\begin{lemma}\label{lem:nsd}
$M(\ol{q}^o(x),\ol{q}^c(x))$ is negative semi-definite if and only if 
\begin{align*}
x\geq \ul{x}\triangleq\max\left\{-\frac{\tq^o-\tq^c}{q^o-q^c},-\frac{\tq^o+(N-1)\tq^c}{q^o+(N-1)q^c}\right\}.
\end{align*}
$M(\ol{q}^o(x),\ol{q}^c(x))$ is negative definite if and only if $x>\ul{x}$.
\end{lemma}

If $x>\ul{x}$, $M(\ol{q}^o(x),\ol{q}^c(x))$ is negative definite, and the optimal best-response is
\begin{align}\label{eq:dual-best-response-many}
    a^*(\omega,x)&=-\frac{1}{2}M^{-1}(\ol{q}^o(x),\ol{q}^c(x))M(\ol{p}^o(x),\ol{p}^c(x))\omega.
\end{align}
The resulting expected value of the dual payoff as a function of $x$ is 
\begin{align}\label{eq:dual_payoff_many}
W(x)&\triangleq\expect[w(a^*(\omega,x),\omega,x,x_0^*)]\\
&=-\frac{1}{4}\expect\left[\omega^T M(\ol{p}^o(x),\ol{p}^c(x))M^{-1}(\ol{q}^o(x),\ol{q}^c(x))M(\ol{p}^o(x),\ol{p}^c(x))\omega\right].
\end{align}

\begin{lemma}\label{lem:dual_payoff_extreme_many}$\lim_{x\downarrow\ul{x}}W(x)=\lim_{x\uparrow+\infty}W(x)=+\infty$, and $\min_{x\geq\ul{x}}W(x)$ exists. 
\end{lemma}
By \autoref{lem:dual_payoff_extreme_many}, the dual payoff admits a minimum at some $x^*>\ul{x}$. We show that this $x^*$ certifies an optimal information structure.
\begin{definition}{\textup{(Linear Disclosure)}}
For $\Lweight=(\lweight_1,\dots,\lweight_N)\in\reals^{N\times N}$, an \emph{$\Lweight-$linear disclosure} is an information structure such that for all $i$ and $\omega$,
\begin{align*}
    s_i=\lweight_i^T\omega=\sum_{j=1}^N \lweight_{i,j}\omega_j.
\end{align*}
\end{definition}
A linear disclosure informs player $i$ about a linear statistic of the state with the responsiveness weights $\lweight_i$. When $N=1$, a linear disclosure corresponds to no disclosure if $\lweight=0$, and to full disclosure otherwise. When $N\geq2$ and $\Lweight$ has full rank, each signal is imperfectly informative, yet the full signal profile perfectly reveals the state. In that case, like a Gaussian coupling, a linear disclosure spreads complete state information across players. When the state is normally distributed, a linear disclosure is a Gaussian information structure. 
\begin{theorem}{\textup{(Optimality of Linear Disclosure)}}\label{thm:lqg-many-linear}
In a normalized symmetric linear-quadratic Gaussian game with a multidimensional state, an $\Lweight^*$-linear disclosure is optimal, where
\begin{align}\label{eq:optimal-linear-many}
    \Lweight^*=-\frac{1}{2}M^{-1}(\ol{q}^o(x^*),\ol{q}^c(x^*))M(\ol{p}^o(x^*),\ol{p}^c(x^*)),
 \end{align}
and $x^*$ is any point in $\argmin_{x> \ul{x}}W(x)$. In the optimal equilibrium, $a_i\equiv s_i$. The induced measure $\pi\in\Delta(A\times\Omega)$ is uniquely optimal.
\end{theorem}
\begin{proof}
By \autoref{lem:dual_payoff_extreme_many}, the dual payoff admits a minimum at some $x^*>\ul{x}$. Consider any such $x^*$.  
By (\ref{eq:dual-best-response-many}), the dual best-response is $a^*(\omega,x^*)=\Lweight^*\omega$; hence, this allocation rule is implementable by incentives, by the contract $\lambda_i(a_i)=-(x^* a_i+x_0^*)$. The allocation rule is symmetric in $i$ and linear in $\omega$. Recall that
\begin{align*}
\udl(x)=\expect[v(a^*(\omega,x),\state)+\sum_{i=1}^{N}(x\action_i^*(\omega,x)+x_0^*)\dot{u}_{i}(a^*(\omega,x),\state)].
\end{align*}
Since $a^*(\omega,x)$ is an interior maximizer and $x$ is an interior minimizer of the dual payoff, it follows by the Envelope Theorem that
\begin{align*}
\left.\frac{d\udl}{d x} \right|_{x=x^*}=\expect\left[\sum_{i=1}^N \action_i^*(\omega,x^*) \dot{u}_{i}(a^*(\omega,x^*),\state)\right]=0.
\end{align*}
By symmetry in $i$, it  follows that for all $i$,
\begin{align}\label{eq:uncorrelated}
&\expect\left[ \action_i^* (\omega,x^*)\dot{u}_{i}(a^*(\omega,x^*),\state)\right]=0.
\end{align}
Because $\action_i^* (\omega,x^*)$ is linear in $\state$  and the components of $\state$ are jointly normally distributed,  $a^*_i(\omega,x^*)$ and $\dot{u}_{i}(a^*(\omega,x^*),\state)$ are also jointly normally distributed. Furthermore, $\expect\left[ \dot{u}_{i}(a^*(\omega,x^*),\state)\right]=0$, and, by (\ref{eq:uncorrelated}), $a^*_i(\omega,x^*)$ and $\dot{u}_{i}(a^*(\omega,x^*),\state)$ are uncorrelated, and thus independent. Therefore, for all $i$,
\begin{align}
&\expect\left[ \dot{u}_{i}(a^*(\omega,x^*),\state)\mid \action_i^*(\omega,x^*)\right]=0\label{eq:zero_cond}.
\end{align}
Therefore, the allocation rule $a^*(\omega,x^*)$ is implementable by information,  by $\Lweight^*$-linear disclosure. By \autoref{thm:dual_certification}, the optimality follows. 

Finally, as the dual agent's best response to $\lambda(x^*,x_0^*)$ is unique, it follows from \autoref{prop:universality} that the corresponding measure is uniquely optimal.
\end{proof}

\autoref{thm:lqg-many-linear} shows that an optimal information structure can be found within a simple class of symmetric noise-free Gaussian information structures (cf. \cite{bergemann2015information}). Moreover, its parameters can be identified by solving a one-dimensional minimization of $W(x)$.

In contrast with the common-state setting of \autoref{sec:lqg-one-dim}, the optimal measure is unique here. Formally, in that setting the optimal certificate leaves the dual agent indifferent among many action profiles,  whereas in this interdependent-value environment it yields a single best response. Conceptually, having one state component per player introduces richer uncertainty, which in turn fully determines the optimal recommendations.

\section{Extensions}\label{sec:extensions}
\paragraph{Bounded Action Spaces}
We have assumed that action spaces are unbounded, to avoid corner solutions and simplify the exposition. However, the certification approach can be easily extended to games with upper and lower bounds on each player's action space, as demonstrated in \autoref{sec:bounded_actions}. The primal problem formulation remains the same, except that the boundary first-order conditions must be inequalities rather than equalities. The dual problem formulation is identical, with the added sign constraints on the contract functions evaluated at the boundary actions. 

\paragraph{Infinite Economies}
Our analysis considers games with a finite number of players. It does not directly cover games with a continuum of players, as considered in the literature on infinite economies, e.g., by \cite{anpa07} and \cite{bemo13a}, even though we anticipate that an analogous certification approach, suitably extended, would work in those settings as well. 

Nonetheless, our analysis enables determining an optimal information structure in a game with a fixed number of players and then examining its limit as the number of players approaches infinity. Our results suggest that in many quadratic economies with a normally distributed state,  Gaussian information structures are certifiably optimal and an optimal aggregate behavior is a deterministic function of a state. With a finite number of players, the latter condition necessitates the designer to either not introduce extraneous noise at all or to ensure that the noises are carefully coupled. With an infinite number of players, this condition can be met by adding independent noises and relying on the law of large numbers; the optimal aggregate behavior can be ensured by the large population size rather than by noise correlation. However,  our analysis suggests that  even in infinite economies, alternative information structures  such as targeted disclosure  may also be optimal, and robustly so.

\paragraph{General Smooth Games}
The tractability of the certification approach in concave games hinges on the capacity of first-order conditions to succinctly represent players' incentives. In general smooth games, i.e., games in which each player's payoff is continuously differentiable in their own action, these first-order conditions may be insufficient, as they can select a suboptimal local maximum or even a minimum in a player's payoff. However, they remain necessary, which enables a straightforward extension  of our analysis: in a smooth but not necessarily concave information-design problem, one  constructs a relaxed primal problem that features only first-order conditions. This problem can be solved using the certification approach. In the final step, one verifies whether players obey the recommendation of the information structure found. If so, this information structure solves the original information-design problem.

\section{Applications}\label{sec:applications}
In this section, we illustrate the developed machinery with three economic applications.\footnote{For a more detailed analysis, see the working version of the paper.}

\subsection{Persuading Investors}\label{sec:investment}
An important question in securities regulation is how much disclosure should be required from firms seeking capital (e.g., \citet*{carvajal2018information}). Although many factors matter, in this section   we focus on two. On the benefit side, greater transparency helps investors reduce risk and make better decisions; on the cost side, it intensifies cream-skimming, as investors herd into the same projects and crowd out investment gains.\footnote{This phenomenon parallels cream-skimming among competing contractors on online platforms, which can likewise be mitigated through information design (\cite{romanyuk2019cream}).} We characterize the optimal balance between these forces within our framework and show that the resulting solution aligns with key venture-capital practices.

Formally, we consider an investment game in the spirit of \cite{anpa07} and \cite{bemo13a}. There are $N\geq 2$ investors, who simultaneously decide how much to invest in a project, $\actions_i=\reals$. The profitability of the project is uncertain; it depends on the unknown project quality $\qual\in\reals$ and on the total amount of investment $\atotal\triangleq\sum_{i=1}^N \action_i$. The ex post payoff  of   player $i$ is
\begin{align}\label{eq:investment_payoff_i}
\ui(\action,\qual)&=(\qual-r \atotal)\action_i-c \action_i,
\end{align}
where $r>0$ is the congestion parameter and $c>0$ is the opportunity cost of investment.
As $r>0$, the project features decreasing returns to scale, i.e., its average profitability decreases in the total investment. We can conveniently rewrite  payoff (\ref{eq:investment_payoff_i}) as 
\begin{align}\label{eq:investment_payoff_i_2}
\ui(\action,\nqual)&=r(\nqual- \atotal)\action_i,
\end{align}
where the state $\nqual\in\states$ is a normalized project quality defined as 
$
\nqual\triangleq (\qual-c)/r.
$

Given  (\ref{eq:investment_payoff_i_2}), for any belief  $\belief\in\Delta(A_{-i}\times\nquals )$, the player $i$'s best response can be found via the first-order condition to equal 
\begin{align}\label{eq:investment_best_response}
a_{i}^{*}(\belief)=\expect_{\belief}\left[\frac{\nqual-\atotali}{2}\right],
\end{align}
where $\atotali\triangleq\sum_{j\neq i}\action_j$, so that the   best response   linearly increases in the normalized quality expectation and linearly decreases in the expected amount of total investment made by other players. The players' actions are thus strategic substitutes.

The information designer has full control over the information regarding the project quality and can privately convey it to each player, thereby persuading that player to invest more or less.  The designer aims to maximize the total profits generated by the project with her ex post payoff being\footnote{Because the expected investment and the expected investment costs are invariant to information, this payoff also captures the objective of maximizing the total investor welfare.}
\begin{align}\label{eq:investment_payoff}
\up(\action,\nqual)=\sum_{i=1}^{N}(\nqual-\atotal)\action_i=(\nqual-\atotal)\atotal=\nqual\atotal-\atotal^2.
\end{align}

\paragraph{Investment Control} If the designer could control the individual investment directly, then she would set the total investment to respond to the state as $\atotal^{FB}(\nqual)=\nqual/2$. As we later show, this first-best allocation rule cannot be implemented via pure information control; however, it can be roughly approximated.
\paragraph{No Disclosure and Full Disclosure}
Two other natural benchmarks are the cases of no disclosure and full disclosure. In these cases, the designer's payoff can be easily computed as:
\begin{align*}
\up^{ND}&=\frac{N}{(N+1)^2}\expect^2[\nqual],\\
\up^{FD}&=\frac{N}{(N+1)^2}(\expect^2[\nqual]+\VAR[\nqual]).
\end{align*}

Comparing the designer's payoffs across the two benchmarks shows that full disclosure strictly dominates no disclosure. However, both information structures suffer from a scaling problem: as the number of players goes to infinity, the designer's payoff and thus the total project profit  converges to zero. In the limit, the individual rent as well as the total profit are dissipated. 
We now show that this problem can be mitigated by adopting an optimal information structure.
\paragraph{Optimal Information}

For any information structure, taking the ex-ante expectation of both sides of (\ref{eq:investment_best_response}), applying the law of iterated expectations, and summing across players (cf. \cite{bergemann2017information}), we obtain that the expected total investment remains constant at
\begin{align*}
\expect[\atotal]=\frac{N}{N+1}\expect[\nqual]=\frac{N}{N+1}\frac{\expect[\qual]-c}{r}>\frac{1}{2}\expect[\nqual]=\expect[\atotal^{FB}(\nqual)].
\end{align*}
Consequently, whenever $\expect[\nqual]>0$, information control can never achieve the full-control benchmark. However, while the designer cannot affect the ex-ante total amount of investment, she can direct investment toward more productive projects. An optimal way to do so is:

\begin{prop}
\textup{\label{prop:investment_info}(Persuading Investment by Exclusivity)} For any number of players, a $1$-targeted disclosure is optimal.
\end{prop}
\begin{proof}
The setting is an instance of the general setting (\ref{eq:common-payoff-gen-1}-\ref{eq:common-payoff-gen-2}) with $\h=N$, $q^o=-1$, $q^c=-N$, $p^o=0$, and $p^c=-N^2$. The formula (\ref{eq:k-opt}) for an optimal $k^*$ becomes:
\begin{align}
k^*=\frac{q^o(\h q^o-p^o)}{q^o(2p^c-\h q^c)-p^oq^c}N=\frac{-1(-N-0)}{-1(-2N^2+N^2)-0}N=\frac{1}{N}N=1,
\end{align}
and thus by \autoref{thm:lqg-common-targeted}, $1$-targeted disclosure is optimal.
\end{proof}
\autoref{prop:investment_info} shows that an optimal information structure takes a very simple form: the designer designates a single player and provides him with exclusive and full access to information. This player makes fully informed decisions. All other players invest the same amount regardless of the project quality. The same information structure is optimal for any number of investors, congestion and cost parameters, and project quality distribution.

The optimal designer's payoff equals 
\begin{align*}  
\up^{*}&=\expect\left[\left(\nqual-\frac{\nqual}{2}-\frac{N-1}{2(N+1)}\expect[\nqual]\right)\left(\frac{\nqual}{2}+\frac{N-1}{2(N+1)}\expect[\nqual]\right)\right]=\frac{N}{(N+1)^2}\expect^2[\nqual]+\frac{1}{4}\VAR[\nqual].
\end{align*}
As such, optimal information design avoids  rent dissipation as the number of players grows to infinity. The project's total profit always  stays above $\VAR[\nqual]/4$   and converges to this level as $N\rightarrow\infty$. The limiting payoff increases with the variance of project quality: under the optimal information structure, just as under full disclosure, riskier projects generate higher realized profits even when their expected quality is the same.\footnote{The setting in this section can also be viewed as Cournot competition with linear production costs and uncertain linear demand. In that interpretation, the rent dissipation under both the no-disclosure and full-disclosure benchmarks parallels the zero-profit outcome of large competitive markets. Exclusive disclosure then amounts to informing only a single firm about the demand state, and this policy maximizes total producer surplus. The structure bears some resemblance to collusion that designates one firm as a monopolist, but information control is weaker than direct production control, because every firm still chooses a positive output even when uninformed. This perspective helps illuminate the role of trade associations, which cannot engage in illegal collusion yet can manage information flows (see, e.g., \citet{kirby1988trade,vives1990trade}).}

Interestingly, this asymmetric treatment of investors mirrors common venture-capital practice: firms cultivate close relationships with a small set of venture capitalists who learn more about the firm than the broader pool of potential investors (e.g., \citet*{bernstein2016impact}). Although several explanations could account for this pattern, our results highlight its informational benefit: by creating a sharp information asymmetry among investors, it curbs congestion and raises both investment profits and investor welfare.

\subsection{Polarizing Predictions}\label{sec:belief_polarization}
Numerous studies  document  societal polarization, with individuals reaching sharply divergent views of the same issues (e.g., \citet*{alesina2020polarization}). Conventional explanations stress behavioral biases or na\"{\i}vet\'e, yet recent evidence shows that most people can reliably distinguish fake news from accurate reporting (\citet*{angelucci2024journalistic}). This tension prompts a sharper question: To what extent can polarization arise among fully rational agents solely because they receive different information, and which information structure generates the greatest polarization?

We address this question in  
a prediction game, in which multiple players try to predict a common underlying state. The state is one-dimensional  $\state\in\reals $ and distributed according to the prior $\prior$. There are $N\geq2$ players, each choosing a prediction $a_i\in A_i=\reals$. The ex post payoff to  player $i$ is
\begin{align*}
\ui(\action,\state)&=-(\action_i-\state)^2.
\end{align*}
Because each player's payoff depends only on his own prediction and the state, for any given belief $\belief\in\Delta(A_{-i}\times\states )$,   player $i$'s best prediction is simply the posterior expectation of the state:
\begin{align*}
a_{i}^{*}(\belief)=\expect_{\belief}[\state].
\end{align*}

We focus on the question of inducing maximal polarization of the players' predictions as measured by the pairwise squared sum:
\begin{align*}
\up(\action,\state)=\sum_{i,j}(\action_i-\action_j)^2.
\end{align*}
Given the payoff, the designer benefits from sending private signals: any public information structure, including full disclosure or no disclosure, leads  the  players to make the same predictions and consequently minimizes the designer's objective. 
An optimal information structure, on the one hand,  must  provide some state information to move players' predictions and, on the other hand, should heterogeneously obfuscate the information to counteract  truth drifting.
\renewcommand{\proofname}{Proof}

\begin{prop}
\textup{\label{cor:polarizing_info_1}(Polarizing by Segregation)} Let the  number of players $N$ be even. Then, an $N/2$-targeted disclosure is optimal.
\end{prop}
\begin{proof}
The setting is an instance of the general setting (\ref{eq:common-payoff-gen-1}-\ref{eq:common-payoff-gen-2}) with $\h=0$, $q^o=-1$, $q^c=0$, $p^o=2N^2$, and $p^c=-2N^2$. The formula (\ref{eq:k-opt}) for an optimal $k^*$ becomes:
\begin{align}
k^*=\frac{q^o(\h q^o-p^o)}{q^o(2p^c-\h q^c)-p^oq^c}N=\frac{-1(-2N^2)}{-1(-4N^2)-0}N=\frac{N}{2},
\end{align}
and thus by \autoref{thm:lqg-common-targeted}, $N/2$-targeted disclosure is optimal.
\end{proof}
\autoref{cor:polarizing_info_1} shows that, irrespective of the prior, expectation polarization can be achieved by a strikingly simple information structure  that informs only half of the population.

This finding aligns with earlier work in information design. In a two-player, binary-state model, \citet{arieli2021feasible} characterized the feasible joint belief distributions and reached a parallel conclusion: to maximize polarization of posterior expectations, it is optimal to inform one of the two players.
\autoref{cor:polarizing_info_1} extends this insight to any number of players and states. \cite{arieli2024population} continue the study of polarization of posterior beliefs. They show that for any even number of players, informing half of them is optimal, because it achieves appropriate statistical bounds. Interestingly, even though belief polarization and expectation polarization differ once the state space is larger than binary, and the underlying arguments are not interchangeable, both problems share the same optimal information structure.

These results contribute to the debate over the sources of societal polarization. Recent studies show that political news consumption is sharply segregated, either because it flows through endogenous peer networks (\cite{bowen2023learning}) or because it is curated by social-media algorithms (\cite{braghieri2024level}). As a result, different groups may learn about different topics. This kind of segregation---learning about different topics, rather than learning different facts about the same topic---can maximally polarize a society. 

\subsection{Informing Competitive Pricing}\label{sec:price_competition}
The information design machinery is useful to understand and guide the design of digital platforms and, more generally, of algorithmic information processing (e.g., \cite{bergemann2024data}, \cite{ichihashi2025buyer}). In this section, we view the designer as a platform that knows demand conditions better than firms do, thanks to a larger, more recent sales dataset and superior analytics. The platform can privately convey this information to each firm by granting it access to personalized data analysis or direct price recommendations, thereby persuading the firm regarding its pricing decisions.  The platform aims to maximize a weighted average of consumer and producer surplus. 
We characterize the information structure  such a designer would optimally design and the resulting price behavior.

Formally, we consider a differentiated duopoly game in the spirit of \cite{vives1984duopoly} and \cite{gal1985information}.\footnote{This formalization complements that adopted by \cite{elliott2022market}, who study information design in unit-demand competition. Naturally, the details of their optimal information structures differ from ours; however, their interpretation of information provision as market segmentation and their broader discussion of how information shapes competitive outcomes can be applied in our setting.} There are two firms that operate in the market. Each firm sells a single product and competes in price with its opponent, so action $\action_i$ is the price set by firm $i$. Demand is ex ante symmetric across firms and is linear in prices and demand shocks:
\begin{align}
\label{eq:consumer_demand}\q_i(\action,\state)&=\state_i-\action_i+\sensni \action_{-i},
\end{align}
where  $\state_i$ captures an individual demand shock for firm $i$'s product and  $\sensni$ is a \emph{cross-price sensitivity} that satisfies $|\sensni|\in(0,1)$, so that $-1<\sensni<0$ corresponds to the case of complementary products, whereas $0<\sensni<1$ corresponds to the case of substitute products. 

The state is two-dimensional $\state=(\state_1,\state_2)\in\reals^2$ and comprises the individual demand shocks. The  shocks  are independently and identically distributed according to a normal distribution, $\state_i\sim N(\mean,\var)$.\footnote{As standard, this specification allows the prices and quantities to be negative.} The firms do not know the state but can be informed about it by the designer. 
The firms have quadratic costs of production so that their profits are
\begin{align}\label{eq:application_payoffs_0}
\ui(\action,\state)&=\action_i\q_i(\action,\state)-c\q_i(\action,\state)^2.
\end{align}
The resulting ex post values of consumer surplus and producer surplus are
\begin{align}\label{eq:application_payoffs_1}
\cs(\action,\state)&=\frac{\action_1^2}{2}+ \frac{\action_2^2}{2}-\sensni\action_1\action_2-\action_1\state_1-\action_2\state_2,\\
\profits(\action,\state)&= u_1(\action,\state)+u_2(\action,\state).
\end{align}
The designer's payoff is a convex combination of  consumer and producer surpluses, with $\delta\in[0,1]$ representing the weight assigned to the consumer surplus:
\begin{align}\label{eq:application_payoffs_3}
\up(\action,\state)=\delta\cs(\action,\state)+(1-\delta)\profits(\action,\state).
\end{align}

The  optimal designer's choices in the extreme cases $\delta=1$ and $\delta=0$ correspond to consumer-optimal and producer-optimal information structures, respectively, whereas the choice in the case $\delta=1/2$ corresponds to the socially efficient information structure. As the welfare weight $\delta$ spans the interval $[0,1]$, the corresponding solutions span the Pareto frontier  in the space of consumer and producer surpluses.

\paragraph{Price Control} We begin the analysis by studying a hypothetical scenario in which the designer can directly control the prices set by the firms. This scenario constitutes a first-best benchmark; it provides an upper bound on the designer's payoff  and illustrates the designer's preferred pricing.

This problem admits a solution only if $\delta$ is not excessively high: there is a threshold value  $\deltaFB\in(0,1)$, such that if $\delta>\deltaFB$, then the designer can arbitrarily increase her payoff by setting arbitrarily large negative prices, because the monetary transfer to consumers outweighs any allocation inefficiency. 
In contrast, if $\delta<\deltaFB$, then the designer's problem is well-behaved; it is concave and admits a unique solution that can be found by first-order  conditions to (\ref{eq:application_payoffs_3}):
\begin{align*}
\action^{FB}_i(\state)=\ri^{FB}\state_i+\rni^{FB}\state_{-i},
\end{align*}
where $\ri^{FB}$ and $\rni^{FB}$ measure the optimal responsiveness of each firm's price to its own demand shock and to the shock of its competitor (explicit formulas appear in \autoref{app:price_control}). 

\paragraph{No Disclosure and Full Disclosure} The designer does not control prices directly but rather indirectly through demand information she supplies to firms. Before deriving the generally optimal policy, it is instructive to analyze two information benchmarks: no disclosure and full disclosure.

Under no disclosure, the firms' beliefs stay at the prior, and the equilibrium prices satisfy the first-order conditions derived from (\ref{eq:application_payoffs_0}).
In equilibrium, each firm sets a price
\begin{align*}
\action^{ND}_i=\frac{1+2c}{2(1+c)-\sensni(1+2c)}\mean.
\end{align*}
Lacking demand information, firms fix their prices at a level proportional to the expected demand. 

In contrast, under full disclosure, the demand shocks are always commonly known. In equilibrium, each firm responds linearly to the shocks perfectly anticipating the price of its opponent:
\begin{align*}
\action^{FD}_i(\state)=\ri^{FD}\state_i+\rni^{FD}\state_{-i}=\frac{2(1+c)(1+2c)}{4(1+c)^2-\sensni^2(1+2c)^2}\state_i+\frac{\sensni(1+2c)^2}{4(1+c)^2-\sensni^2(1+2c)^2}\state_{-i}.
\end{align*}
In a sense, this behavior enriches   price-setting under no disclosure. If $\state_1=\state_2=\mean$, then prices are the same as those under no disclosure, $\action^{FD}_i(\mean,\mean)=\action^{ND}_i$. If $\state_1\neq\state_2$, then   demand is asymmetric across firms, and   prices are adjusted to reflect competitive advantages. However, the prices average to the prices under no disclosure, $\expect[\action^{FD}_i(\state)]=\action^{ND}_i$.

\paragraph{Optimal Information}
The choice of any of the extreme information structures has drawbacks. Providing no disclosure misses the opportunity to strengthen the link between demand and allocation and thus potentially limits efficiency. Providing full disclosure  may exacerbate competition and dissipate firm profits. Instead, and as a direct consequence of \autoref{thm:lqg-many-linear},  the optimal information structure is partially informative  and takes a  form of a linear disclosure.

\begin{prop}\label{prop:optimal_demand_info}\textup{(Managing Competition by Linear Statistics)} There is a parameter $\hat\delta(\eta,c)$ such that if $\delta\neq\hat\delta(\eta,c)$, an optimal direct information structure exists, is unique, and recommends for some coefficients $a_0^*$, $\ri^*$ and $\rni^*$  prices
\begin{align}\label{eq:optimal-competition}
a_i(\state)=a_0^*+\ri^* \state_i+\rni^* \state_{-i}.
\end{align}
\end{prop}
The uniquely optimal direct information structure treats the firms symmetrically. 
Under it, each firm can only infer a linear combination of the individual demand shocks, and thus not able to infer whether the recommendation of a given price stems from its own demand conditions or the conditions of its competitor. Generically, $\rni^*\neq\rni^{FD}$ and $\ri^*\neq\ri^{FD}$, so providing full disclosure is suboptimal; instead, each firm receives personalized information that differs from its competitor's. This finding suggests that restricting disclosure to be public, as often done in the oligopoly literature on information sharing (\cite{vives1990trade,vives1999oligopoly}), though natural in some contexts, may entail losses.

\autoref{prop:optimal_demand_info} enables the calculation of optimal information structures. It can be shown that the optimal information structure may exhibit a discontinuous change with respect to the consumer weight $\delta$ at $\hat\delta(\eta,c)$, accompanied by a change in equilibrium price volatility and correlations.\footnote{Formally, while the optimal certificate changes continuously for all $\delta\in(0,1)$, at $\delta=\deltacr$ the certificate allows a range of the dual agent's best responses connecting one branch of best responses to another, so that the implemented allocation rule displays a jump.} Intuitively, the pricing induced under the optimal information structure seeks to approximate its first-best counterpart under direct price control. However, lacking the ability to enforce prices directly, the designer chooses an information regime that better approximates it.

This observation yields two  insights. First, patterns of price volatility and correlations can serve as effective diagnostics for discerning which side of the market, consumers or producers, the information structure is designed to favor. Second, even a slight shift in the design objective can trigger a dramatic change in the information provision and in resulting market behavior. This, in turn, can serve as a cautionary tale, underscoring the potential for market instability introduced by algorithms, which may pursue shifting objectives and adapt their policies more rapidly than human decision makers.

\section{Conclusion}\label{sec:conclusion}
In this paper, we introduced a certification approach to solving information-design problems in games and demonstrated its effectiveness and tractability in symmetric linear-quadratic settings. In doing so, we provided theoretical justification for the use of targeted disclosure, Gaussian coupling, and linear disclosure. Our findings shed light on disclosure practices that guide investment, on the socially efficient control of information in markets, and on the limits of Bayesian polarization. 

Our analysis lays the groundwork and offers tools for studying information design in general smooth games. We see at least three promising avenues for further research. First, our theoretical framework could be applied to other important economic settings, such as contests, public-good provision, and labor or financial markets. Second, one could develop more general sufficient conditions for strong duality in non-compact settings. Third, the framework could be expanded to incorporate elements like information elicitation, information spillovers across players, and dynamic interaction. We expect all of these extensions to be feasible within the approach we have outlined.

\bibliographystyle{ecta}
\bibliography{general}
\newgeometry{tmargin=1.1in,bmargin=1.1in,lmargin=0.9in,rmargin=0.9in,footskip=0.4in}
\appendix

\section{Appendix}\label{app:A}
\subsection{Formalism Omitted in \autoref{sec:model}}
\paragraph{Derivation of First-Order Condition (\ref{eq:foc_belief})}
For $\varepsilon>0$, define
\begin{align*}
    \phi_{i \varepsilon}^+(a_{i},a_{-i},\state)&\triangleq\frac {u_{i}(a_{i}+\varepsilon,a_{-i},\state)-u_{i}(a_{i},a_{-i},\state)}{\varepsilon},\\
    \phi_{i \varepsilon}^-(a_{i},a_{-i},\state)&\triangleq\frac {u_{i}(a_{i},a_{-i},\state)-u_{i}(a_{i}-\varepsilon,a_{-i},\state)}{\varepsilon}.
\end{align*}

\begin{definition} \textup{(Admissible Equilibria)}
An equilibrium is admissible, if for any player $i$ and equilibrium belief $\belief_i\in\Delta(A_{-i}\times\states)$, there exists a best response $a_{i}^{*}\in A_i$, $\Delta>0$, and $\psi:A_{-i}\times \states\to \mathbb{R}$ such that $\int_{A_{-i}\times \states} \psi(a_{-i},\state)d\mu_i<+\infty$ and for all $\varepsilon\in(0,\Delta)$, $(\action_{-i},\state)\in A_{-i}\times\states$, $|\phi_{i \varepsilon}^+(a_i^*,a_{-i},\state)|\leq \psi(a_{-i},\state)$ and $|\phi_{i \varepsilon}^-(a_i^*,a_{-i},\state)|\leq \psi(a_{-i},\state)$ .
\end{definition}

This admissibility effectively requires players' payoff differences to be locally well behaved at the best responses. It ensures the interchangeability of operators in (\ref{eq:foc_belief}) and is satisfied in all equilibria that we characterize. (If $\actions$ and $\states$ were compact, then admissibility would be trivially satisfied whenever each ${u}_{i}(a,\state)$ was continuously differentiable in $a_i$. For general $\actions$ and $\states$, any equilibrium is admissible if, for instance, each $u_i(a,\omega)$ is polynomial.)

\begin{lemma}\textup{(First-Order Approach)}
Under \autoref{ass:concavity}, for any player $i$ and admissible equilibrium belief $\belief_i\in\Delta(A_{-i}\times\states)$,  $a_{i}$ is player $i$'s best response if and only if condition (\ref{eq:foc_belief}) is satisfied.
\end{lemma}
\begin{proof}
For any  $i$ and admissible equilibrium belief  $\belief_i$, $a_{i}^{*}$  satisfies the following condition:
\begin{align}\label{eq:foc_br}
\int_{\actions_{-i}\times\states} u_i(a_{i}^*,a_{-i},\state)\textrm{d} \belief_i \geq \int_{\actions_{-i}\times\states} u_i(a_{i},a_{-i},\state)\textrm{d} \belief_i,\quad \forall a_i\in A_i.
\end{align}
Under \autoref{ass:concavity}, condition (\ref{eq:foc_br}) is satisfied if and only if   for any $\varepsilon>0$, the following holds: 
\begin{align}
\int_{\actions_{-i}\times\states} \phi_{i \varepsilon}^+(a_{i}^*,a_{-i},\state)\textrm{d} \belief_i\leq 0,\label{eq:phi_pos}\\
\int_{\actions_{-i}\times\states} \phi_{i \varepsilon}^-(a_{i}^*,a_{-i},\state)\textrm{d} \belief_i\geq 0.\label{eq:phi_neg}
\end{align}
Moreover, $\phi_{i \varepsilon}^+$ is decreasing in $\varepsilon$ and $\phi_{i \varepsilon}^-$ is increasing in $\varepsilon$; thus,  conditions (\ref{eq:phi_pos}), (\ref{eq:phi_neg}) are equivalent to
\begin{align*}
0\geq\lim_{\varepsilon\rightarrow 0}\int_{\actions_{-i}\times\states} \phi_{i \varepsilon}^+(a_{i}^*,a_{-i},\state)\textrm{d} \belief_i
&=\lim_{\varepsilon\rightarrow 0}\frac{1}{\varepsilon}\int_{\actions_{-i}\times\states} u_{i}(a_{i}^*+\varepsilon,a_{-i},\state)-u_{i}(a_{i}^*,a_{-i},\state)\ \textrm{d} \belief_i\\
&=\left.\frac{\partial^{+}}{\partial a_i}\int_{\actions_{-i}\times\states} u_{i}(a_{i},a_{-i},\state)\textrm{d} \belief_i \right|_{a_i=a_i^{*}},
\end{align*}
\begin{align*}
0\leq\lim_{\varepsilon\rightarrow 0}\int_{\actions_{-i}\times\states} \phi_{i \varepsilon}^-(a_{i}^*,a_{-i},\state)\textrm{d} \belief_i
&=\lim_{\varepsilon\rightarrow 0}\frac{1}{\varepsilon}\int_{\actions_{-i}\times\states}  u_{i}(a_{i}^*,a_{-i},\state)-u_{i}(a_{i}^*-\varepsilon,a_{-i},\state)\ \textrm{d} \belief_i,\\
&=\left.\frac{\partial^{-}}{\partial a_i}\int_{\actions_{-i}\times\states} u_{i}(a_{i},a_{-i},\state)\textrm{d} \belief_i \right|_{a_i=a_i^{*}} .
\end{align*}
By Lebesgue's dominated convergence theorem (Theorem 11.32, \cite{rudin1976principles}), because  belief $\belief_i$ arises in an admissible equilibrium, the following holds:
\begin{align*}
\lim_{\varepsilon\rightarrow 0}\int_{\actions_{-i}\times\states} \phi_{i \varepsilon}^+(a_{i}^*,a_{-i},\state)\textrm{d}\belief_i&=\int_{\actions_{-i}\times\states} \lim_{\varepsilon\rightarrow 0}\phi_{i \varepsilon}^+(a_{i}^*,a_{-i},\state)\textrm{d}\belief_i=\int_{\actions_{-i}\times\states} \dot u_{i}(a_{i}^*,a_{-i},\state) \textrm{d}\belief_i,\\
\lim_{\varepsilon\rightarrow 0}\int_{\actions_{-i}\times\states} \phi_{i \varepsilon}^-(a_{i}^*,a_{-i},\state)\textrm{d}\belief_i&=\int_{\actions_{-i}\times\states} \lim_{\varepsilon\rightarrow 0}\phi_{i \varepsilon}^-(a_{i}^*,a_{-i},\state)\textrm{d}\belief_i=\int_{\actions_{-i}\times\states} \dot u_{i}(a_{i}^*,a_{-i},\state) \textrm{d}\belief_i,
\end{align*}
and thus 
\begin{align*}
  \left.\frac{\partial}{\partial a_i}\int_{\actions_{-i}\times\states} u_{i}(a_{i},a_{-i},\state)\textrm{d} \belief_i \right|_{a_i=a_i^{*}}=\int_{\actions_{-i}\times\states} \dot u_{i}(a_{i}^*,a_{-i},\state) \textrm{d}\belief_i=0.
\end{align*}
\end{proof}
\subsection{Formalism Omitted in \autoref{sec:general_analysis}}
\paragraph{Proof of \autoref{thm:dual_certification}} 

The first step is
\begin{lemma}
\label{thm:weak_duality}\textup{(Weak Duality)} 
$
V^P\leq V^D.
$
\end{lemma}
\begin{proof}
Take any dual variables $(\lambda,\gamma)$ that satisfy  the constraints of the dual problem (\ref{problem:sender_dual}).
Take any  measure $\pi$ that satisfies the constraints of the primal problem (\ref{problem:sender_primal}). Integrating both sides of the dual constraints over $a\in A$   and $\state\in \states$ against measure $\pi$
yields:
\begin{alignat}{1}
\int_{A\times\states}v(a,\state)\textrm{d}\pi & \leq\int_{A\times\states}\sum_{i=1}^{N}\lambda_{i}(a_{i})\dot{u}_{i}(a,\state)\textrm{d}\pi+\int_{A\times\states}\gamma(\state)\textrm{d}\pi=\int_{\states}\gamma(\state)\textrm{d}\mu_{0}\label{eq:weak_duality_derivation},
\end{alignat}
where the equality follows because $\pi$ satisfies the primal constraints. The left-hand side of (\ref{eq:weak_duality_derivation}) is  the value of the primal problem given measure $\pi$. The right-hand side of (\ref{eq:weak_duality_derivation}) is the value of the dual problem given dual variables $(\lambda,\gamma)$. As the inequality (\ref{eq:weak_duality_derivation}) holds for any allowed values of primal measure and dual variables, it also holds 
 at the respective maximization and minimization limits.
\end{proof}

Continuing with the proof of \autoref{thm:dual_certification},
 take any primal measure $\pi$ implementable by information, i.e., that satisfies the constraints of the primal problem (\ref{problem:sender_primal}). If it is implementable by incentives,  then there exist dual variables $\lambda$ that implement this measure in the dual problem (\ref{problem:adversarial_contracting}), and 
\begin{alignat*}{1}
\vdual & =\inf_{\lambda^{\prime}\in\times_{i}\meas(A_{i})} \expect_{\mu_0}\left[\sup_{a\in A}\ud(a,\state,\lambda^{\prime})\right]\\
& \leq \expect_{\pi}\left[\ud(a,\state,\lambda)\right] \\
& = \int_{A\times\states}v(a,\state)\textrm{d}\pi - \int_{A\times\states}\sum_{i=1}^{N}\lambda_{i}(a_{i})\dot{u}_{i}(a,\state)\textrm{d}\pi\\
& = \int_{A\times\states}v(a,\state)\textrm{d}\pi \leq \vprimal,
\end{alignat*} 
where the first inequality follows from the implementability of $\pi$ in the dual problem and the last three steps follow from the feasibility of $\pi$ in the primal problem. 

Furthermore, by \autoref{thm:weak_duality}, $\vdual\geq\vprimal$. Combining the two inequalities, we obtain 
\begin{alignat*}{1}
\vdual  = \int_{A\times\states}v(a,\state)\textrm{d}\pi = \vprimal,
\end{alignat*}
which proves the optimality of measure $\pi$.

\paragraph{Proof of \autoref{prop:universality}} For any optimal measure $\pi$, the following holds: 
\begin{gather*}
    \expect_{\pi}[v(a,\state)]=V^P,\\
    \int_{A'_{i}\times A_{-i}\times\states}\dot{u}_{i}(a,\state)\textrm{d}\pi=0\quad\forall\,i=1,\dots,N,\textrm{measurable\ }A'_{i}\subseteq A_{i}.
\end{gather*}
As such, if the dual agent is offered contract $\lambda$ and plays according to an optimal measure then his expected payoff is
\begin{align*}
   \expect_{\pi} [\ud(a,\state,\lambda)]&= \expect_{\pi}[ v(a,\state)-\sum_{i=1}^{N}\lambda_{i}(a_{i})\dot{u}_{i}(a,\state)]\\
   &=\expect_{\pi}[v(a,\state)]-\expect_{\pi}[\sum_{i=1}^{N}\lambda_{i}(a_{i})\dot{u}_{i}(a,\state)]=V^P.
\end{align*}
That is, the dual agent obtains the same expected payoff of $V^P$ by playing according to an  optimal measure, whether $\pi$ or $\pi'$.

Moreover, if $\lambda$ certifies the optimality of measure $\pi$, then playing according to $\pi$ is a best response of the dual agent to contract $\lambda$. However, since, as shown above, $\pi'$ delivers the same expected dual payoff as $\pi$, it follows that $\pi'$ is also a best response of the dual agent to contract $\lambda$. 
Indeed, for every $\state\in\states$ and $a\in A$, $\ud(a,\state,\lambda)\leq \sup_{\tilde a\in A}\ud(\tilde a,\state,\lambda)$.
Since $\lambda$ certifies $\pi$, we have $V^{D}=V^{P}$, and since $\pi'$ is optimal, $\expect_{\pi'}[\ud(a,\state,\lambda)]=V^{P}$. Hence
\begin{align*}
0=V^{D}-\expect_{\pi'}[\ud(a,\state,\lambda)]=\int_{\states}\left(\sup_{a\in A}\ud(a,\state,\lambda)-\int_{A}\ud(a,\state,\lambda)\,\pi'(\mathrm{d}a\mid \state)\right)\mu_{0}(\mathrm{d}\state).
\end{align*}
The integrand is nonnegative, so it must be equal to zero for $\mu_{0}$-almost every $\state$. Therefore, for $\mu_{0}$-almost every $\state$, the measure $\pi'(\cdot\mid\state)$ is supported on $\arg\max_{a\in A}\ud(a,\state,\lambda)$, that is, $\pi'$ is also implementable by incentives under contract $\lambda$.

Finally, as an optimal measure, $\pi'$ satisfies the constraints of the primal problem. Hence, $\pi'$ is implementable by information and by incentives; by definition, $\lambda$ certifies the optimality of $\pi'$.

\paragraph{Proof of \autoref{thm:strong_duality}}
Assume that $\Omega$ and all $A_i$ are compact (metric) spaces, and that $v$ and each $\dot{u}_i$ are continuous on $X \triangleq \Omega\times A$. We first introduce an auxiliary dual problem in which the multipliers are restricted to be continuous:
\begin{align*}
V^{D,c}\triangleq \inf_{\ga\in C(\Omega),\ \lambda_i\in C(A_i)}
&\int_{\Omega}\ga(\omega)\,d\mu_0\\
\text{s.t. }\quad
&\ga(\omega)+\sum_{i=1}^N \lambda_i(a_i)\dot{u}_i(a,\omega)\geq v(a,\omega)
\quad \forall (a,\omega)\in A\times\Omega.
\end{align*}
Because every continuous function is measurable, the feasible set of $V^{D,c}$ is contained in the feasible set of $V^D$. Hence $V^D\leq V^{D,c}$. By weak duality, $V^P\leq V^D$. Therefore it is enough to prove that $V^{D,c}=V^P$.

Define $E\triangleq C(X)$, $Y\triangleq C(\Omega)\times \prod_{i=1}^N C(A_i)$, each endowed with the sup norm. Since $X$ is compact metric, both are Banach spaces, and by the Riesz-Markov theorem  \cite[Theorem 6.19]{Rudin1987}
$E^*=M(X)$, $Y^*=M(\Omega)\times \prod_{i=1}^N M(A_i)$,
where $M(\cdot)$ denotes the finite signed Radon measures.

Define the continuous linear operator $T:Y\to E$ by
\begin{align*}
T(\ga,\lambda)(a,\omega)\triangleq \ga(\omega)+\sum_{i=1}^N \lambda_i(a_i)\dot{u}_i(a,\omega).
\end{align*}
Also define the proper convex functions $\phi:E\to\mathbb{R}\cup\{+\infty\}$ and $\psi:Y\to\mathbb{R}$ by
\begin{align*}
\phi(f)\triangleq
\barray
0, &\text{if } f(x)\geq v(x)\ \forall x\in X,\\
+\infty, &\text{otherwise},
\earray\quad
\psi(\ga,\lambda)\triangleq \int_{\Omega}\ga(\omega)\,d\mu_0.
\end{align*}
Then,
\begin{align*}
V^{D,c}=\inf_{y\in Y}\Big\{\psi(y)+\phi(Ty)\Big\}.
\end{align*}

Now choose $M>\|v\|_{\infty}$ and let $y_0=(M,0,\ldots,0)\in Y$. Then $Ty_0\equiv M>v$ on $X$, so $\phi$ is continuous at $Ty_0$. Hence the Banach-space Fenchel-Rockafellar theorem applies and yields \cite[Theorems 1 and 3]{Rockafellar1967}
\begin{align*}
V^{D,c}
=
\sup_{\pi\in E^*}
\Big\{
-\psi^*(T^*\pi)-\phi^*(-\pi)
\Big\}.
\end{align*}

We now compute the two conjugates. First,
\begin{align*}
-\phi^*(-\pi)
&=
-\sup_{f\in E}
\left\{
\int_X f\,d(-\pi)-\phi(f)
\right\}\\
&=
\inf_{f\in C(X),\ f\geq v}\int_X f\,d\pi.
\end{align*}
Therefore
\begin{align*}
-\phi^*(-\pi)
=
\barray
\int_X v\,d\pi, &\text{if } \pi\in M_+(X),\\
-\infty, &\text{otherwise}.
\earray
\end{align*}
Indeed, if $\pi\in M_+(X)$, the infimum is attained at $f=v$. If $\pi\notin M_+(X)$, then by the Riesz-Markov theorem there exists $h\in C(X)$ with $h\geq 0$ and $\int_X h\,d\pi<0$; hence for $f_n\triangleq v+nh$ we have $f_n\geq v$ and
\begin{align*}
\int_X f_n\,d\pi
=
\int_X v\,d\pi+n\int_X h\,d\pi
\to -\infty.
\end{align*}

Second, identify $T^*\pi\in Y^*$ explicitly. For $\pi\in M(X)$, let $\pi_0\in M(\Omega)$ be the $\Omega$-marginal of $\pi$, and for each $i$ define a finite signed measure $\nu_i\in M(A_i)$ by
\begin{align*}
\nu_i(B)\triangleq \int_{B\times A_{-i}\times \Omega}\dot{u}_i(a,\omega)\,d\pi
\qquad \forall \text{ Borel } B\subseteq A_i.
\end{align*}
Since $\dot{u}_i$ is continuous on the compact set $X$, it is bounded, so $\nu_i$ is well defined. Moreover, for every $(\ga,\lambda)\in Y$,
\begin{align*}
\langle T^*\pi,(\ga,\lambda)\rangle
&=
\int_X \ga(\omega)\,d\pi
+\sum_{i=1}^N\int_X \lambda_i(a_i)\dot{u}_i(a,\omega)\,d\pi\\
&=
\int_{\Omega}\ga\,d\pi_0+\sum_{i=1}^N\int_{A_i}\lambda_i\,d\nu_i.
\end{align*}
Hence, under the identification $Y^*=M(\Omega)\times \prod_{i=1}^N M(A_i)$,
\begin{align*}
T^*\pi=(\pi_0,\nu_1,\ldots,\nu_N).
\end{align*}

Now $\psi(\ga,\lambda)=\int_{\Omega}\ga\,d\mu_0$ depends only on $\ga$, so its conjugate is
\begin{align*}
\psi^*(\eta_0,\eta_1,\ldots,\eta_N)
=
\barray
0, &\text{if } \eta_0=\mu_0 \text{ and } \eta_i=0 \ \forall i,\\
+\infty, &\text{otherwise}.
\earray
\end{align*}
Therefore
\begin{align*}
-\psi^*(T^*\pi)
=
\barray
0, &\text{if } \pi_0=\mu_0 \text{ and } \nu_i=0 \ \forall i,\\
-\infty, &\text{otherwise}.
\earray
\end{align*}

Combining the two conjugate calculations gives
\begin{align*}
V^{D,c}
=
\sup_{\pi\in M_+(X)}
\left\{
\int_X v\,d\pi
\ \middle|\
\pi_0=\mu_0,\ \nu_i=0\ \forall i
\right\}.
\end{align*}
Since $\pi_0=\mu_0$ and $\mu_0$ is a probability measure, every such $\pi$ has total mass one, so $\pi\in \Delta(X)$.

Finally, the condition $\pi_0=\mu_0$ is exactly the Bayes-plausibility constraint. Also, $\nu_i=0$ is equivalent, by the uniqueness part of the Riesz-Markov theorem, to
\begin{align*}
\int_{A_i'\times A_{-i}\times \Omega}\dot{u}_i(a,\omega)\,d\pi=0
\qquad
\forall \text{ Borel } A_i'\subseteq A_i,
\end{align*}
which is exactly the obedience constraint in the primal problem. Hence the feasible set above is precisely the primal feasible set, and therefore $V^{D,c}=V^P$. We conclude that
\begin{align*}
V^P\leq V^D\leq V^{D,c}=V^P,
\end{align*}
so $V^D=V^P$.

\subsection{Formalism Omitted in \autoref{sec:lqg-one-dim}} \label{app:section4}
\paragraph{Game Normalization}
The game normalization is achieved by defining $\omega'=\omega-\expect[\omega]$, $a_i'=a_i+\frac{\expect[\omega]+\bias_1}{q^o+q^c}$, $b=\h\expect[\omega]+\bias_2-2(p^o+p^c)\frac{\expect[\omega]+\bias_1}{q^o+q^c}$, and ignoring the strategically irrelevant additive terms, i.e., those in $u_i$ that do not depend on $a_i$ and those in $v$ that do not depend on $a$.

\paragraph{Proof of \autoref{thm:lqg-common-targeted}, continued.}\label{app:section4_targeted}
If $k^*\notin[0,N]$, then, since $q^o p^c>p^o q^c$, either (i)   $p^o<\h q^o$, 
or (ii)   $(\h q^o-p^o)(q^o+q^c)>2(p^c q^o-p^o q^c)$.

Indeed, if neither (i) or (ii) holds then it must be the case that  $p^o\geq \h q^o$, and $(\h q^o-p^o)(q^o+q^c)\leq 2(p^c q^o-p^o q^c)$. But in this case, because of our assumptions $q^o<0$, $q^o(\h q^o-p^o)\geq 0$ and therefore,
\begin{align*}
\frac{k^*}{N}=\frac{q^o(\h q^o-p^o)}{(2(p^c q^o-p^o q^c)-(\h q^o-p^o)(q^o+q^c))+q^o(\h q^o-p^o)}\in[0,1].
\end{align*}
Consequently, if $k^*\notin[0,N]$, then either (i) or (ii) holds.

\textbf{Case (i).} If (i) holds, then  no disclosure is optimal, which can be certified by an affine contract with $(x,x_0)=(-\h,-\frac{\bias}{q^o+q^c})$. Under this contract, the dual payoff function is 
\begin{align*}
    w(a,\omega)=x_0\omega+(p^o-\h q^o)\avsq+(p^c-\h q^c)\av^2=x_0\omega+\frac{1}{N^2}a^T M((p^o-\h q^o)N+p^c-\h q^c,p^c-\h q^c)a,
\end{align*}
where $M(x,y)$ denotes the $N\times N$ matrix whose diagonal entries equal $x$ and off-diagonal entries equal $y$. By \autoref{thm:dual_certification}, no disclosure is optimal if $a=(0,\dots,0)$ maximizes $w(a,\omega)$ for all $\omega$. This happens if $M((p^o-\h q^o)N+p^c-\h q^c,p^c-\h q^c)$ is negative semidefinite (NSD). Note that matrix $M(x,y)$ has two distinct eigenvalues: $x+(N-1)y$ and $x-y$. Therefore, $M(x,y)$ is NSD if $x+(N-1)y\leq 0$ and $x-y\leq 0$ which for $M((p^o-\h q^o)N+p^c-\h q^c,p^c-\h q^c)$ corresponds to:
\begin{align*}
    p^o&\leq \h q^o,\\
    p^o+p^c&\leq \h (q^o+q^c).
\end{align*}

Subcase 1: $\h q^c\geq p^c$. In this case, NSD is equivalent to $p^o-\h q^o\leq 0$. In case $q^o p^c>p^oq^c$ and $p^o<\h q^o$, this is satisfied. 

Subcase 2: $\h q^c<p^c$. In this case, NSD is equivalent to $p^o-\h q^o+p^c-\h q^c\leq 0$. 
In case $q^o p^c>p^oq^c$ and $p^o<\h q^o$, note that $p^c=-p^o-\eta$ and $q^c=-q^o-\e$ for some $\eta$ and $\e>0$. Because $q^op^c>p^oq^c$, we have $-q^o\eta>-p^o\e$, which implies $\eta/\e>p^o/q^o>h$ because $p^o<\h q^o<0$. Thus, $\eta>\h\e$; equivalently, $p^o+p^c<\h(q^o+q^c)$.

\textbf{Case (ii).} If (ii) holds, then  full disclosure is optimal, which can be certified by an affine contract with $(x,x_0)=(\h-\frac{2(p^o+p^c)}{q^o+q^c},-\frac{\bias}{q^o+q^c})$. Under this contract, the dual payoff function is 
\begin{align*}
w(a,\omega)=&x_0\omega+\left(2\h-\frac{2(p^o+p^c)}{q^o+q^c}\right)\omega \av+\left(p^o+q^o(\h-\frac{2(p^o+p^c)}{q^o+q^c})\right) \avsq +\left(p^c+q^c(\h-\frac{2(p^o+p^c)}{q^o+q^c})\right) \av^2\\
=&x_0\omega+\frac{1}{N^2}\left(
(2\h-\frac{2(p^o+p^c)}{q^o+q^c})N\omega 1_N^Ta+a^T M(x,y)a
\right),
\end{align*}
where $1_N=(1,\ldots,1)^T$ is an $N$-dimensional vector of ones, $x=N(p^o+q^o(\h-\frac{2(p^o+p^c)}{q^o+q^c}))+p^c+q^c(\h-\frac{2(p^o+p^c)}{q^o+q^c})$ and $y=p^c+q^c(\h-\frac{2(p^o+p^c)}{q^o+q^c})$.

By \autoref{thm:dual_certification},  full disclosure  is optimal if $a=(-\frac{\omega}{q^o+q^c},\ldots,-\frac{\omega}{q^o+q^c})$ maximizes $w(a,\omega)$ for all $\omega$. If $M(x,y)$ is NSD, then the maximizing $a$ can be found via F.O.C.:
\begin{align*}
\left(2\h-\frac{2(p^o+p^c)}{q^o+q^c}\right)N\omega 1_N+2M(x,y)a=0,
\end{align*}
with a solution $a=(-\frac{\omega}{q^o+q^c},\ldots,-\frac{\omega}{q^o+q^c})$. Therefore, it remains to establish that $M(x,y)$ is NSD, $x+(N-1)y\leq 0$ and $x-y\leq 0$ which corresponds to  
\begin{align*}
p^o+q^o\left(\h-\frac{2(p^o+p^c)}{q^o+q^c}\right)&\leq 0,  &&  \\
p^o+q^o\left(\h-\frac{2(p^o+p^c)}{q^o+q^c}\right)+p^c+q^c\left(\h-\frac{2(p^o+p^c)}{q^o+q^c}\right)&\leq 0.&&
\end{align*}

Subcase 1: $p^c+q^c(\h-\frac{2(p^o+p^c)}{q^o+q^c})\leq 0$ and $p^o+p^c\neq h(q^o+q^c)$. We show that the inequalities hold strictly. It suffices to show $p^o+q^o(\h-\frac{2(p^o+p^c)}{q^o+q^c})<0$; equivalently, $(\h q^o-p^o)(q^o+q^c)>2(p^cq^o-p^oq^c)$. In case $p^cq^o >p^oq^c$ and $(\h q^o-p^o)(q^o+q^c)>2(p^cq^o-p^oq^c)$, this is clearly satisfied. 

Subcase 2: $p^c+q^c(\h-\frac{2(p^o+p^c)}{q^o+q^c})>0$ and $p^o+p^c\neq h(q^o+q^c)$. We show that the inequalities hold strictly. It suffices to show $p^o+q^o(\h-\frac{2(p^o+p^c)}{q^o+q^c})+p^c+q^c(\h-\frac{2(p^o+p^c)}{q^o+q^c})<0$; equivalently, $\h(q^o+q^c)-p^o-p^c<0$. 
In case $q^o p^c-p^oq^c>0$ and $(\h q^o-p^o)(q^o+q^c)>2(p^cq^o-p^oq^c)$, noting that $p^c=-p^o-\eta$ and $q^c=-q^o-\e$ for some $\eta\in\reals$ and $\e>0$, we have $-q^o\eta>-p^o\e$ and $q^o(2\eta-\h\e)>p^o\e$, implying (by summing them side-by-side) $q^o(\eta-\h\e)>0$; equivalently, $\eta-\h\e<0$. Thus,
\begin{align*}
\h(q^o+q^c)-p^o-p^c=\eta-\h\e<0.
\end{align*}

\paragraph{Proof of \autoref{thm:lqg-common-gaussian}, continued.}\label{thm:lqg-common-gaussian-app}
By concavity conditions $-q^o>0$, and thus $\sigma^2>0$ if and only if 
\begin{align*}
  \beta(1+q^c\beta+q^o\beta)&>0,&&\\
  \frac{(p^o-\h q^o)}{2(p^c q^o-p^o q^c)}\left(1+\frac{(q^c+q^o)(p^o-\h q^o)}{2(p^c q^o-p^o q^c)}\right)&>0,&&\\
  \frac{(p^o-\h q^o)(2(p^c q^o-p^o q^c)+(q^c+q^o)(p^o-\h q^o))}{4(p^c q^o-p^o q^c)^2}&>0.&&
\end{align*}
Because $q^o<0$, setting $c_1\triangleq q^o(\h q^o-p^o)$  and $c_2\triangleq q^o(2p^c-\h q^c)-p^o q^c$, the inequality holds if and only if 
\begin{align*}
  q^o(\h q^o-p^o)(2(p^c q^o-p^o q^c)+(q^c+q^o)(p^o-\h q^o))&>0,&&\\
  c_1(c_2-c_1)&>0.&&
\end{align*}
Moreover,
\begin{align*}
    \frac{k^*}{N}=\frac{q^o(\h q^o-p^o)}{q^o(2p^c-\h q^c)-p^o q^c}=\frac{c_1}{c_2}.
\end{align*}
The result follows, because for any $c_1,c_2\in\reals$, $c_1/c_2\in(0,1)$ if and only if $c_1(c_2-c_1)>0$.

\subsection{Formalism Omitted in \autoref{sec:investment}} 
\paragraph{No Disclosure} Under no disclosure, each player's action cannot depend on the state and is thus uniquely determined by condition 
\begin{align} 
\expect[a_{i}]=\frac{1}{N+1}\expect[\nqual].
\end{align}
\paragraph{Full Disclosure} For any $\nqual$, the ensuing game admits a strictly concave potential $\Psi(a,\nqual)=(\nqual-\frac{\atotal}{2})\atotal-\frac{1}{2}\sum_{i=1}^N a_i^2$ and thus has a unique equilibrium. Parameterize a symmetric linear strategy profile: 
\begin{align*}
     \action_i(\nqual)=k_0+k_1 \nqual.
\end{align*}
The best-response condition (\ref{eq:investment_best_response}) can  be rewritten as
\begin{align*}
     \action_i(\nqual)=-\frac{N-1}{2}k_0+\frac{1-k_1(N-1)}{2} \nqual,
\end{align*}
which determines the equilibrium parameters at $k_0=0$ and $k_1=\frac{1}{N+1}$.

\subsection{Formalism Omitted in \autoref{sec:lqg-multi-dim}}
\paragraph{Game Normalization}\label{sec:lgq-normalization}
Under any information structure, the expected marginal payoffs must equal  zero. Thus,
\begin{align*}
    \expect[\omega_i p^o+\tstateni p^c+\bias_1+q^o a_i + q^c \tactionni ]=\mu(p^o+(N-1)p^c)+\bias_1+\expect[q^o a_i+q^c \tactionni ]=0.
\end{align*}
By symmetry, it follows that for all $i$,
\begin{align*}
    \expect[a_i]=a_0\triangleq-\frac{\mu(p^o+(N-1)p^c)+\bias_1}{q^o+(N-1)q^c}.
\end{align*}
In the normalized game, the expected actions equal zero. The normalization is achieved by defining $\omega'=(\omega-\mu)/\sigma$, $a_i'=(a_i-a_0)/\sigma$, $b=\bigl(b_2 +(\tilde p^o+(N-1)\tilde p^{c})\mu + 2\bigl(\tilde q^o + (N-1)\tilde q^c\bigr)a_0\bigr)/\sigma$, rescaling the payoffs by $\sigma^2$, and ignoring the strategically irrelevant additive terms, i.e., those in $u_i$ that do not depend on $a_i$ and those in $v$ that do not depend on $a$.

\paragraph{Proof of \autoref{lem:nsd}}
Matrix $M(x,y)$ has two distinct eigenvalues: $x+(N-1)y$ and $x-y$. Therefore, it is negative semidefinite if and only if $x+(N-1)y\leq 0$ and $x-y\leq 0$; it is negative definite if and only if $x+(N-1)y<0$ and $x-y<0$. Therefore, $M(\ol{q}^o,\ol{q}^c)=M(\tq^o+x q^o,\tq^c+x q^c)$ is negative semidefinite if and only if 
\begin{align*}
    x(q^o+(N-1)q^c)&\leq-\tq^o-(N-1)\tq^c,\\
    x(q^o-q^c)&\leq -(\tq^o-\tq^c).
\end{align*}
By concavity assumptions, $q^o+(N-1)q^c<0$ and $q^o-q^c<0$. The result follows.

\paragraph{Proof of \autoref{lem:dual_payoff_extreme_many}}
Note that $M(x,y)M(z,w)=M(xz+(N-1)yw,xw+yz+(N-2)yw)$ and $M(x,y)^{-1}=M(\frac{x+(N-2)y}{(x-y)(x+(N-1)y)},-\frac{y}{(x-y)(x+(N-1)y)})$, if invertible. Thus,
\begin{align*}
    a^*(\omega,x)&=-\frac{1}{2}M^{-1}(\ol{q}^o(x),\ol{q}^c(x))M(\ol{p}^o(x),\ol{p}^c(x))\omega\\
&=-\frac{1}{2}\frac{M(\ol{q}^o(x) \ol{p}^o(x)+(N-2)\ol{q}^c(x) \ol{p}^o(x)-(N-1)\ol{q}^c(x)\ol{p}^c(x),\ol{q}^o(x) \ol{p}^c(x)-\ol{q}^c(x)\ol{p}^o(x))}{(\ol{q}^o(x)-\ol{q}^c(x))(\ol{q}^o(x)+(N-1)\ol{q}^c(x))}\omega,
\end{align*}
and
\begin{align*}
W(x)&=-\frac{1}{4}\expect\left[\omega^T M(\ol{p}^o(x),\ol{p}^c(x))M^{-1}(\ol{q}^o(x),\ol{q}^c(x))M(\ol{p}^o(x),\ol{p}^c(x))\omega\right]\\
&=-\frac{N}{4}\frac{Z(x)}{(\ol{q}^o(x)-\ol{q}^c(x))(\ol{q}^o(x)+(N-1)\ol{q}^c(x))},
\end{align*}
where 
\begin{align*}
Z(x)\triangleq\ol{q}^o(x)&\Bigl(
\ol{p}^o(x)^2+\ol{p}^c(x)^2(N-1)(1+(N-2)\rho)+2(N-1)\rho\ol{p}^o(x)\ol{p}^c(x)\Bigr)+\\
&+\ol{q}^c(x)\Bigl(
\ol{p}^o(x)^2(N-2-\rho(N-1))-\rho(N-1)^2\ol{p}^c(x)^2-2(N-1)\ol{p}^o(x)\ol{p}^c(x)\Bigr).
\end{align*}

First, consider the case of $x\downarrow \ul{x}$. In this case, the term dividing $Z(x)$ is going to zero (from above). Thus, if $Z(\ul{x})\neq 0$, then as $x\downarrow \ul{x}$, $W(x)$ diverges to $+\infty$.

(i) $\ul{x}=-\frac{\tq^o-\tq^c}{q^o-q^c}$. As $-\frac{\tq^o-\tq^c}{q^o-q^c}\geq -\frac{\tq^o+(N-1)\tq^c}{q^o+(N-1)q^c}$, we have $\tq^o q^c\leq  q^o\tq^c$, and by the genericity assumption, $\tq^o q^c< q^o\tq^c$.
This implies that $\ol{q}^o=\ol{q}^c<0$. Thus,
\begin{align*}
Z(\ul{x})=\ol{q}^c(N-1)(1-\rho)(\ol{p}^o-\ol{p}^c)^2<0.
\end{align*}

(ii) $\ul{x}=-\frac{\tq^o+(N-1)\tq^c}{q^o+(N-1)q^c}$. As $-\frac{\tq^o-\tq^c}{q^o-q^c}\leq -\frac{\tq^o+(N-1)\tq^c}{q^o+(N-1)q^c}$, we have $\tq^o q^c\geq  q^o\tq^c$, and by the genericity assumption, $\tq^o q^c> q^o\tq^c$.
This implies that $\ol{q}^o+(N-1)\ol{q}^c=0$, and $\ol{q}^c>0$. Thus,
\begin{align*}
Z(\ul{x})=-\ol{q}^c(1+\rho(N-1))(\ol{p}^o+(N-1)\ol{p}^c)^2<0.
\end{align*}

Now, consider the case of $x\uparrow+\infty$. In this case, for the denominator of the dual payoff, we have 
\begin{align*}
\frac{(\ol{q}^o-\ol{q}^c)(\ol{q}^o+(N-1)\ol{q}^c)}{x^2}\to (q^o-q^c)(q^o+(N-1)q^c)>0,
\end{align*}
while for the numerator we have 
\begin{align}
\frac{Z(x)}{x^3}\to&\ 
q^o\Bigl(
(p^o)^2+(p^c)^2(N-1)(1+(N-2)\rho)+2(N-1)\rho p^o p^c\Bigr)\notag\\
&+q^c\Bigl(
(p^o)^2(N-2-\rho(N-1))-\rho(N-1)^2(p^c)^2-2(N-1)p^op^c\Bigr)\label{eq:proof_extreme_dual}.
\end{align}

It suffices to show that the right-hand side of (\ref{eq:proof_extreme_dual}) is strictly negative, because then the dual payoff is $x$ times a strictly positive constant, which diverges to infinity as $x$ grows.

First, note that 
\begin{align*}
&(p^o)^2+(p^c)^2(N-1)(1+(N-2)\rho)+2(N-1)\rho p^o p^c\\
&=(p^o+\rho(N-1) p^c)^2+(p^c)^2((N-1)(1+(N-2)\rho)-\rho^2(N-1)^2)\\
&>(p^c)^2(N-1)(1+(N-2)\rho-\rho^2(N-1))\\
&=(p^c)^2(N-1)(1+(N-1)\rho)(1-\rho)\geq 0.
\end{align*}

(i) $q^c<0$. Because $q^o<q^c$, 
the right-hand side of (\ref{eq:proof_extreme_dual}) is lower than
\begin{align*}
&q^c\Bigl(
(p^o)^2+(p^c)^2(N-1)(1+(N-2)\rho)+2(N-1)\rho p^o p^c\Bigr)\\
&+q^c\Bigl(
(p^o)^2(N-2-\rho(N-1))-\rho(N-1)^2(p^c)^2-2(N-1)p^op^c\Bigr)\\
&=q^c(p^o-p^c)^2(N-1)(1-\rho)<0.
\end{align*}

(ii) $q^c>0$. Because $q^o+(N-1)q^c<0$, the right-hand side of (\ref{eq:proof_extreme_dual}) is lower than
\begin{align*}
&-(N-1)q^c\Bigl(
(p^o)^2+(p^c)^2(N-1)(1+(N-2)\rho)+2(N-1)\rho p^o p^c\Bigr)\\
&+q^c\Bigl(
(p^o)^2(N-2-\rho(N-1))-\rho(N-1)^2(p^c)^2-2(N-1)p^op^c\Bigr)\\
&=-q^c(p^o+(N-1)p^c)^2(1+\rho(N-1))<0.
\end{align*}

The result follows.
\subsection{Formalism Omitted in \autoref{sec:price_competition}}\label{app:price_competition}
We can microfound the linear demand as being generated by a continuum of consumers that differ in their tastes. Each consumer has a type $\state=(\state_1,\state_2)\in\reals^2$ and decides how much of the firms' products to consume, $\q=(\q_1,\q_2)$. The payoff of a type-$\state$ consumer who consumes quantities $\q$ at prices $\action$ is
\begin{align*}
-\action^T \q+\frac{1}{2}(\state-\q)^T \uconsmat^{-1} (\state-\q),
\end{align*}
where $\uconsmat$ is an $N\times N$ negative semidefinite matrix:
\begin{align*}
\uconsmat\triangleq
\left(
\begin{array}{cc}
-1 & \sensni \\
\sensni & -1
\end{array}
\right).
\end{align*}
For any price vector $\action\in \actions$, the quantities demanded by a type-$\state$ consumer are $\q(\action,\state)=\state+\uconsmat\action$.

\noindent Consumer and producer surpluses can be written as
\begin{align*}
\cs(\action,\state)&=\action^T\Bp_{CS}\state-\frac{1}{2}\action^T \Cp_{CS}\action,\\
\profits(\action,\state)&=\action^T(\state+\uconsmat\action)-c(\state+\uconsmat\action)^T(\state+\uconsmat\action)=-c\state^T\state+\action^T\Bp_{PS}\state-\frac{1}{2}\action^T\Cp_{PS}\action,
\end{align*}
where the payoff coefficient matrices are
\begin{align*}
&\Bp_{CS}\triangleq-I,\ \Cp_{CS}\triangleq\uconsmat,\\
&\Bp_{PS}\triangleq I-2c\uconsmat,\  \Cp_{PS}\triangleq-2\uconsmat+2c\uconsmat^2,\\
&\Bp\triangleq\delta \Bp_{CS}+(1-\delta)\Bp_{PS},\ \Cp\triangleq\delta \Cp_{CS}+(1-\delta)\Cp_{PS}.
\end{align*}
 
\paragraph{Price Control}\label{app:price_control} The designer's first-order condition is
\begin{align*}
\Bp\state-\Cp\action=0,
\end{align*}
which results in the first-best responsiveness matrix
$\resp^{FB}=\Cp^{-1}\Bp$, and thus,
\begin{align*}
    \ri^{FB}&=\frac{
  2 + \delta(6\delta-7)
  + 4c^2(1 - \delta)^2(1 - \eta^2)
  + 2c(1 -\delta)(3 - 5\delta - (1 - \delta)\eta^2)
}{  (1 - \eta^2)
  \bigl(2 - 3\delta + 2c(1 - \delta)(1 - \eta)\bigr)
  \bigl(2 - 3\delta + 2c(1 - \delta)(1 + \eta)\bigr)
},\\
\rni^{FB}&=\frac{
  \eta(
      2 + \delta(6\delta-7)
      + 4c^2(1 - \delta)^2(1 - \eta^2)
      + 4c(1 - \delta)(1 - 2\delta)
  )
}{ (1 - \eta^2)
  \bigl(2 - 3\delta + 2c(1 - \delta)(1 - \eta)\bigr)
  \bigl(2 - 3\delta + 2c(1 - \delta)(1 + \eta)\bigr).
}
\end{align*}
The threshold value $\deltaFB$ is the one that equalizes the determinant of $\Cp$ to zero:
\begin{align*}
\deltaFB=\frac{2+2c(1-|\sensni|)}{3+2c(1-|\sensni|)}.
\end{align*}

\paragraph{No Disclosure and Full Disclosure} Equilibrium pricing behavior is derived from the system of first-order conditions:
\begin{align*}
\expect_{\mu}[\q_i(\action_i,\action_{-i},\state)+\frac{\partial \q_i(\action_i,\action_{-i},\state)}{\partial \action_i}(\action_i-2c q_i(\action_i,\action_{-i},\state))]=0,\quad i=1,2.
\end{align*}
\paragraph{Proof of \autoref{prop:optimal_demand_info}}
The setting is an instance of the general framework (\ref{eq:payoffs-general-many-1}-\ref{eq:payoffs-general-many-2}) with $N=2$, $\rho=0$, $b_1=b_2=0$ and $\po=1+2c$, $\pc=0$, $\qo=-2(1+c)$, $\qc=\sensni(1+2c)$, $\pop=\delta(-1)+(1-\delta)(1+2c)$, $\pcp=(1-\delta)(-2c\sensni)$, $\qop=\delta/2 +(1-\delta)(-1-c(1+\sensni^2))$, $\qcp=-\delta\sensni/2+(1-\delta)\sensni(1+2c)$.

The non-genericity conditions (\ref{eq:many_gener_1}-\ref{eq:many_gener_3}) reduce to
\begin{align}\label{eq:critical_value}
    \delta\neq\deltacr(\sensni,c)\triangleq\frac{2 + 6 c + 4 c^2 - 8 c |\sensni| - 8 c^2 |\sensni| + 2 c \sensni^2 + 4 c^2 \sensni^2}{5 + 8 c + 4 c^2 - |\sensni| - 10 c |\sensni| - 8 c^2 |\sensni| + 2 c \sensni^2 + 4 c^2 \sensni^2}.
\end{align}
If $\sensni<0$, at $\delta=\deltacr$ condition (\ref{eq:many_gener_2}) is violated. If $\sensni>0$, at $\delta=\deltacr$ condition (\ref{eq:many_gener_3}) is violated.  Otherwise, the conditions hold.

The resulting lower bound on $x$ is
     \begin{align}\label{eq:xl}
        \xl=\max\left\{-\frac{\qop-\qcp}{\qo-\qc},-\frac{\qop+\qcp}{\qo+\qc}\right\}.
     \end{align}
The  expected dual payoff reduces to
\begin{align}
    \udl(x)=-\frac{\var(\qol(x)\pol(x)^2+\qol(x)\pcl(x)^2-2\qcl(x)\pol(x)\pcl(x))}{2(\qol(x)^2-\qcl(x)^2)}.\label{eq:dual_competition}
\end{align}
By \autoref{thm:lqg-many-linear} and normalization (\ref{sec:lgq-normalization}), whenever  $\delta\neq\deltacr$, the optimal direct information structure is unique and is given by (\ref{eq:optimal-competition}) with 
\begin{align*}
    \ri^*=-\frac{\qol(x^*)\pol(x^*)-\qcl(x^*)\pcl(x^*)}{2(\qol(x^*)^2-\qcl(x^*)^2)},\\
    \rni^*=-\frac{\qol(x^*)\pcl(x^*)-\qcl(x^*)\pol(x^*)}{2(\qol(x^*)^2-\qcl(x^*)^2)},
\end{align*}
and  $x^*$ being any minimizer of (\ref{eq:dual_competition}) over $x^*>\xl$. The constant term $a_0^*$ can be found via mean action invariance: $a_0^*=a_i^{ND}-(\ri^*+\rni^*)\mean.$

\section{Bounded Action Spaces}\label{sec:bounded_actions}

Consider the concave information-design problem as in the main text, but let the action space of each player be $\actions_i=[\amin_i,\amax_i]$, $-\infty<\amin_i<\amax_i<+\infty$.\footnote{The extension to half-bounded spaces is straightforward.} For any given $\belief\in\Delta(A_{-i}\times\states )$, the player's best-response action  $a_{i}^{*}(\belief)$, if interior, must be unimprovable by local deviations to lower and higher actions and hence satisfies the  first-order condition
\begin{gather*}
\mathbb{E}_{\belief}\left[\dot{u}_{i}(a_{i}^{*},a_{-i},\state)\right]=0.
\end{gather*}
In contrast, the optimal boundary actions must only be unimprovable by  one-sided local deviations. As such, the player's best response if located on the boundary must satisfy the following:
\begin{gather}
\mathbb{E}_{\belief}\left[\dot{u}_{i}(\amin_{i}^{*},a_{-i},\state)\right]\leq0,\\
\mathbb{E}_{\belief}\left[\dot{u}_{i}(\amax_{i}^{*},a_{-i},\state)\right]\geq0.
\end{gather}
We can write the resulting primal information-design problem as follows:
\begin{alignat}{1}
\vprimalb\triangleq\sup_{\pi\in\Delta(A\times\states)} & \int_{A\times\states}v(a,\state)\textrm{d}\pi\label{problem:sender_primal_bdd}\\
\textrm{s.t.\ } & \int_{A'_{i}\times A_{-i}\times\states}\dot{u}_{i}(a,\state)\textrm{d}\pi=0\quad\forall\,i=1,\dots,N,\textrm{measurable\ }A'_{i}\subseteq (\amin_i,\amax_i),\label{eq:obedience_constraints_bdd_1}\\
& \int_{A'_{i}\times A_{-i}\times\states}\dot{u}_{i}(a,\state)\textrm{d}\pi\leq0\quad\forall\,i=1,\dots,N,\textrm{measurable\ }A'_{i}\subseteq [\amin_i,\amax_i),\label{eq:obedience_constraints_bdd_2}\\
& \int_{A'_{i}\times A_{-i}\times\states}\dot{u}_{i}(a,\state)\textrm{d}\pi\geq0\quad\forall\,i=1,\dots,N,\textrm{measurable\ }A'_{i}\subseteq (\amin_i,\amax_i],\label{eq:obedience_constraints_bdd_3}\\
 & \int_{A\times\states'}\textrm{d}\pi=\int_{\states'}\textrm{d}\mu_{0}\quad\forall\,\textrm{measurable\ } \states'\subseteq\states.\label{eq:feasibility_constraints_bdd}
\end{alignat}
This primal problem entails the  dual problem 
\begin{alignat}{1}
\vdualb\triangleq\inf_{\lambda\in\times_{i}\meas(A_{i}),\gamma\in\meas(\states)} & \ \int_{\states}\gamma(\state)\textrm{d}\mu_{0}\label{problem:sender_dual_bdd}\\
\textrm{s.t.}\ \sum_{i=1}^{N} & \lambda_{i}(a_{i})\dot{u}_{i}(a,\state)+\gamma(\state)\geq v(a,\state)\ \forall\,a\in A,\state\in\states,\\
& \lambda_i(\amin_i)\geq0,\ \lambda_i(\amax_i)\leq0\quad\forall\,i=1,\dots,N\label{eq:sign_constraints_primal_bdd}.
\end{alignat}
The presence of additional obedience constraints (\ref{eq:obedience_constraints_bdd_2}), (\ref{eq:obedience_constraints_bdd_3})  in the primal problem translates into the sign constraints on the  Lagrange multipliers (\ref{eq:sign_constraints_primal_bdd}) in the dual problem. As in the case of unbounded actions, the dual problem (\ref{problem:sender_dual_bdd})  can be simplified and rewritten as
\begin{align}\label{problem:adversarial_contracting_bdd}
\vdualb=&\inf_{\lambda\in\times_{i}\meas(A_{i})} \expect_{\mu_0}[\sup_{a\in A}\ud(a,\state,\lambda)]
\\
\textrm{s.t.\ } & \lambda_i(\amin_i)\geq0,\ \lambda_i(\amax_i)\leq0\quad\forall\,i=1,\dots,N\label{eq:sign_constraints_dual_bdd}.
\end{align}
The adversarial-contracting interpretation remains intact, but the space of allowed contracts is limited at the boundary actions by the presence of sign constraints.

\begin{lemma}
\label{thm:weak_duality_bdd}\textup{(Weak Duality with Bounded Action Spaces)} 
$\label{eq:weak_duality_bdd}
\vprimalb\leq \vdualb.
$
\end{lemma}
\begin{proof}
Take any dual variables $(\lambda,\gamma)$ that satisfy  the constraints of the dual problem (\ref{problem:sender_dual_bdd}).
Take any  measure $\pi$ that satisfies the constraints of  primal problem (\ref{problem:sender_primal_bdd}). Integrating both sides of the dual constraints over $a\in A$   and $\state\in \states$ against  measure $\pi$
yields
\begin{alignat}{1}
\int_{A\times\states}v(a,\state)\textrm{d}\pi & \leq\int_{A\times\states}\sum_{i=1}^{N}\lambda_{i}(a_{i})\dot{u}_{i}(a,\state)\textrm{d}\pi+\int_{A\times\states}\gamma(\state)\textrm{d}\pi\leq\int_{\states}\gamma(\state)\textrm{d}\mu_{0}\label{eq:weak_duality_derivation_bdd},
\end{alignat}
where the second inequality follows because $\pi$ satisfies the primal constraints and the Lagrange multipliers satisfy the dual constraints. (This inequality holds as equality in the case of unbounded actions.) The left-hand side of (\ref{eq:weak_duality_derivation_bdd}) is  the value of the primal problem given measure $\pi$, whereas the right-hand side of (\ref{eq:weak_duality_derivation_bdd}) is the value of the dual problem given dual variables $(\lambda,\gamma)$. As  inequality (\ref{eq:weak_duality_derivation_bdd}) holds for any allowed values of primal measure and dual variables, it also  holds  at the respective maximization and minimization limits.
\end{proof}

In the case of bounded actions, we call a measure $\pi\in\Delta(A\times\states)$  \emph{implementable by information} if it satisfies the constraints of the primal problem (\ref{problem:sender_primal_bdd}) and \emph{implementable by incentives} if there exists a feasible contract in the dual problem (\ref{problem:adversarial_contracting_bdd}) that induces this measure as a best response and  satisfies the complementarity slackness condition: for all $i$ and measurable $A_i'\subseteq[\underline{a}_i,\overline{a}_i]$, $\int_{A'_{i}\times A_{-i}\times\states}\lambda_i(a_i)\dot{u}_{i}(a,\state)\textrm{d}\pi=0$.

\begin{theorem}
\textup{(Optimality Certification with Bounded Action Spaces)} \label{thm:dual_certification_bdd} 
In the case of bounded actions,  if $\pi\in\Delta(A\times\states)$ is implementable by information and is implementable by incentives by contract $\lambda$, then (i) $\pi$ solves the information-design problem, (ii) $\lambda$ solves the adversarial-contracting problem, and (iii) $\vprimalb=\vdualb$.
\end{theorem}

\begin{proof}
Take any primal measure $\pi$ implementable by information, i.e., one that satisfies the constraints of the primal problem (\ref{problem:sender_primal_bdd}). If it is implementable by incentives,  then there exist dual variables $\lambda$ that implement this measure in the dual problem (\ref{problem:adversarial_contracting_bdd}) and satisfy complementarity slackness. Therefore, 
\begin{alignat*}{1}
\vdualb & =\inf_{\lambda^{\prime}\in\times_{i}\meas(A_{i})+(\ref{eq:sign_constraints_dual_bdd})} \expect_{\mu_0}\left[\sup_{a\in A}\ud(a,\state,\lambda^{\prime})\right]\\
& \leq \expect_{\pi}\left[\ud(a,\state,\lambda)\right] \\
& = \int_{A\times\states}v(a,\state)\textrm{d}\pi - \int_{A\times\states}\sum_{i=1}^{N}\lambda_{i}(a_{i})\dot{u}_{i}(a,\state)\textrm{d}\pi\\
& = \int_{A\times\states}v(a,\state)\textrm{d}\pi \leq \vprimalb,
\end{alignat*} 
where the first inequality follows from the implementability of $\pi$ in the dual problem, whereas the last three steps follow from the feasibility of $\pi$ in the primal problem and  complementarity slackness.

Furthermore, by \autoref{thm:weak_duality_bdd}, $\vdualb\geq\vprimalb$. Hence,
\begin{alignat*}{1}
\vdualb  = \int_{A\times\states}v(a,\state)\textrm{d}\pi = \vprimalb,
\end{alignat*}
which proves the optimality of measure $\pi$.
\end{proof}
\begin{claim}
\textup{(Strong Duality with Bounded Action Spaces)} \label{thm:strong_duality-bounded} If $\Omega$ is compact, and $v$ and each $\dot{u}_i$ are continuous, then $\vdualb=\vprimalb$.
\end{claim}
\begin{proof}
The proof is analogous to that of \autoref{thm:strong_duality}, except that one carries the boundary sign restrictions through the auxiliary continuous dual. In particular, one defines $V_B^{D,c}$ exactly as in the proof of \autoref{thm:strong_duality}, but with continuous multipliers $\lambda_i\in C(A_i)$ constrained to satisfy $\lambda_i(\underline a_i)\geq 0$ and $\lambda_i(\bar a_i)\leq 0$ for each $i$. Then $V_B^D\leq V_B^{D,c}$, while weak duality gives $V_B^P\leq V_B^D$, so it suffices to show $V_B^{D,c}=V_B^P$. Applying the Fenchel-Rockafellar duality with the same spaces and linear operator as in the proof of \autoref{thm:strong_duality}, the only change is that the functional on the multiplier space now includes the indicators of the cones $K_i=\{\lambda_i\in C(A_i):\lambda_i(\underline a_i)\geq 0,\ \lambda_i(\bar a_i)\leq 0\}$. Accordingly, the conjugate condition in the dual becomes $\pi_0=\mu_0$ together with $\nu_i\in K_i^\circ$ for every $i$, where $\nu_i$ is the weighted marginal measure induced by $\dot u_i$ and $K_i^\circ$ is the polar cone of $K_i$. A direct characterization of $K_i^\circ$ shows that this is equivalent to the bounded-action obedience constraints \eqref{eq:obedience_constraints_bdd_1}--\eqref{eq:obedience_constraints_bdd_3}: zero weighted mass on interior subsets and the appropriate weak inequalities on sets touching the lower or upper boundary. Thus the feasible set in the Fenchel-Rockafellar dual coincides exactly with the primal feasible set, so $V_B^{D,c}=V_B^P$, and therefore $V_B^P\leq V_B^D\leq V_B^{D,c}=V_B^P$, implying $V_B^D=V_B^P$.

\end{proof}

\section{Online Appendix}\label{sec:online_app}

In this appendix, we apply the certification approach to a game with  non-quadratic payoffs.

Imagine a designer who can inform a representative retailer and a representative consumer about a new two-sided application. The application is either high quality ($\omega = 1$) or low quality ($\omega = 0$).

The retailer chooses a search intensity $a_{1}\in[0,1]$ in the application, and the  consumer chooses $a_{2}\in[0,1]$. Each player $i$ incurs the cost
\begin{align*}
c(a_i)=
\begin{cases}
\hat{a}^{3}a_i, & a_i\in[0,\hat{a}),\\
\frac{a_i^{4}}{4}+\frac{3\hat{a}^{4}}{4}, & a_i\in[\hat{a},1].
\end{cases}
\end{align*}
where  $\hat{a}=3/4$. Thus $c$ is smooth and convex, and $\hat{a}$ minimizes the average-cost function.

The players' payoffs are
\begin{align*}
u_i(a,\omega)=\omega\,a_i a_{-i}-c(a_i).
\end{align*}

A higher search intensity $a_i$ raises player $i$'s chance of finding a match, and this benefit is greater when the state is high and when the other player searches more intensively. Hence the game exhibits strategic complementarities.

The designer aims to raise overall search intensity to boost engagement and advertising revenue. Her payoff is
\begin{align*}
v(a,\omega)=a_{1}+a_{2}.
\end{align*}

\paragraph{Public Information}

First, suppose the platform sends a public signal, identical for both players. The optimal signal can then be derived via the standard concavification procedure. Let $\mu$ denote the common belief that $\omega=1$. Given this belief, the players play either (i) $a_1=a_2=a^{*}(\mu)\ge\hat{a}$, where
\begin{align*}
\mu\,a^{*}(\mu) = \bigl(a^{*}(\mu)\bigr)^{3},
\end{align*}
so $a^{*}(\mu)=\sqrt{\mu}\ge\hat{a}$; or (ii) $a_1=a_2=0$. Consequently, the designer's indirect utility equals $2\sqrt{\mu}$ whenever $\sqrt{\mu}\ge\hat{a}$ and $0$ otherwise.

Hence, if the prior satisfies $\mu_{0}\ge\hat{a}^{2}$, no disclosure is optimal. If $\mu_{0}<\hat{a}^{2}$, the optimal public signal splits posteriors between $0$ and $\hat{a}^{2}$; concretely,
(i) when $\omega = 0$, the posterior is either $0$ or $\hat{a}^{2}$, leading to actions $(a_{1},a_{2})=(0,0)$ or $(\hat{a},\hat{a})$;
(ii) when $\omega = 1$, the posterior is always $\hat{a}^{2}$, so $(a_{1},a_{2})=(\hat{a},\hat{a})$.

\paragraph{Optimal Information}

Now, we show that  the above public information is optimal even within the class of all private information structures.

The dual payoff is  
\beas
w(a,\omega)=a_1+a_2-\lambda_1(a_1)(\omega a_2-c'(a_1))-\lambda_2(a_2)(\omega a_1-c'(a_2)).
\eeas

Posit the following certificate:
\beas
\lambda_i(a_i)=\barray
-\frac{a_i}{\hat{a}^3}, \andif a_i<\hat{a},\\
-\frac{1}{a_i^2}, \andif a_i\geq \hat{a},
\earray
\eeas
so that
\beas
w(a,\omega)=\barray
\omega(\frac{a_2}{a_1^2}+\frac{a_1}{a_2^2}), \andif a_1,a_2\geq \hat{a},\\
\omega(\frac{a_2}{a_1^2}+\frac{a_1a_2}{\hat{a}^3}), \andif a_1\geq \hat{a}> a_2,\\
\omega(\frac{a_1}{a_2^2}+\frac{a_1a_2}{\hat{a}^3}), \andif a_2\geq \hat{a}> a_1,\\
\omega(\frac{2a_1a_2}{\hat{a}^3}), \andif a_1,a_2< \hat{a}.
\earray
\eeas

From the perspective of the dual agent, if $\omega=0$, then any  $a$ is optimal, including $a=(0,0)$ and $(\hat{a},\hat{a})$. If $\omega=1$,  we can focus on the  case $a_1,a_2\geq \hat{a}$. Because $w(a,\omega)$ is convex on this domain, its maximum must occur at one of the corner points $a=(\hat{a},\hat{a})$, $(\hat{a},1)$, $(1,\hat{a})$, or $(1,1)$. A straightforward calculation shows that $(\hat{a},\hat{a})$ is optimal whenever $1-\hat{a}-\hat{a}^{2}\le 0$, which indeed holds for $\hat{a}=3/4$.
 The result follows. 
\end{document}